\documentclass[twocolumn,floatfix,amssymb,amsmath,secnumarabic,nofootinbib, nobalancelastpage]{revtex4-1}    

\usepackage[pdftex]{graphicx} 
\usepackage{dcolumn}  
\usepackage{bm}       
\usepackage[usenames,dvipsnames]{color}
\definecolor{URLCOL}{rgb}{0,0.52,0.83} 
\definecolor{LINKCOL}{rgb}{0.05,0.5,0} 
\definecolor{CITECOL}{rgb}{0.25,0,0.48} 
\usepackage{epstopdf}
\usepackage[pdftex,bookmarks,breaklinks,bookmarksopen,bookmarksnumbered,colorlinks,
                  linkcolor=LINKCOL,linktocpage,citecolor=CITECOL,urlcolor=URLCOL,
                     pdfpagemode=UseOutline,pdftex,pagebackref]{hyperref}
\usepackage{datetime}

\def\preprintlink{ \href{http://dft.uci.edu + PAPER REF}{title of paper} }
\def\preprinttext{~}

\usepackage{fancyhdr}
\makeatletter
\fancypagestyle{titlepage}
{
       \lhead{\textsc{\preprinttext\ ~ \href{http://dft.uci.edu/publications.php}{~}}}
       \chead{}
       \rhead{\preprintlink}
       \lfoot{}
}
\makeatother
\def\preprintlink{ 
       \href{http://dft.uci.edu}
        {
~}
       }

\usepackage{graphicx}
\usepackage{amsmath}
\usepackage{bm}
\usepackage{threeparttable}
\usepackage[normalem]{ulem}
\usepackage{array,multirow}
\usepackage{appendix}
\usepackage{epstopdf}
\usepackage{booktabs}
\usepackage{natbib}
\usepackage{feynmp}
\usepackage{threeparttable}
\definecolor{TITLECOL}{rgb}{0.1,0.2,0.7} 
\definecolor{PCOL}{rgb}{0.5,0.06,0.01} 
\definecolor{CHAPCOL}{rgb}{0,0.48,0} 
\definecolor{SECOL}{rgb}{0.1,0.2,0.7} 
\definecolor{CONTENTSCOL}{rgb}{0.1,0.2,0.7} 
\definecolor{SSECOL}{rgb}{0.25,0,0.48} 
\definecolor{SSSECOL}{rgb}{0.2,0.08,0.53} 
\definecolor{SHDCOL}{rgb}{0.4,0,0} 
\definecolor{ITMCOL}{rgb}{0.4,0,0} 
\definecolor{EXCOL}{rgb}{0,0.47,0.01} 
\definecolor{DEFCOL}{rgb}{0,0.42,0.01} 

\def\coloredtitle#1{\title{\textcolor{TITLECOL}{#1}}} 

\definecolor{URLCOL}{rgb}{0,0.17,0.43} 
\definecolor{LINKCOL}{rgb}{0.05,0.4,0} 
\definecolor{CITECOL}{rgb}{0.35,0,0.48} 

\definecolor{ngreen}{rgb}{0,0.48,0}

\def\sec#1{\section{\textcolor{SECOL}{#1}}}
\def\ssec#1{\subsection{\textcolor{SSECOL}{#1}}}
\def\sssec#1{\subsubsection{\textcolor{SSSECOL}{#1}}}

\def\sectable#1{
\addcontentsline{toc}{subsection}{~~Table: \textcolor{SSECOL}{#1}}
\begin{table}[h]
\caption{\bf \textcolor{SSECOL}{#1}}
}

\def\bea{\begin{eqnarray}}
\def\eea{\end{eqnarray}}
\def\ben{\begin{equation}}
\def\een{\end{equation}}
\def\benu{\begin{enumerate}}
\def\enu{\end{enumerate}}

\def\bei{\begin{itemize}}
\def\eei{\end{itemize}}
\def\beit{\begin{itemize}}
\def\eit{\end{itemize}}
\def\benu{\begin{enumerate}}
\def\enu{\end{enumerate}}

\def\n{n}

\def\sss{\scriptscriptstyle\rm}

\def\1var{(\bx_1...\bx\N)}

\def\half{\frac{1}{2}}

\def\br{{\bf r}}

\def\bx{{x}}

\def\bj{{\bf j}}

\def\x{_{\sss X}}

\def\s{_{\sss S}}
\def\xc{_{\sss XC}}

\def\N{_{\sss N}}

\def\HO{_{\sss HO}}

\def\LDA{^{\rm LDA}}

\def\BA{^{\rm BA}}

\def\LBA{^{\rm LBA}}

\def\GEA{^{\rm GEA}}
\def\GE{^{\rm GE}}
\def\W{^{\rm W}}
\def\GGA{^{\rm GGA}}

\def\B88{^{\rm B88}}

\def\TF{^{\rm TF}}

\def\sph_int{ {\int d^3 r}}

\definecolor{SPECOL}{rgb}{0,0.47,0.01}
\definecolor{QUOCOL}{rgb}{0,0,0.2}
\definecolor{SHDCOLb}{rgb}{0.69,0.4,0.1}

\definecolor{SPEQ}{rgb}{0.01,0.4,0.05} %

\definecolor{SPEQv}{rgb}{0.45,0.05,0.45} %

\definecolor{SPEQb}{rgb}{0.01,0.1,0.65} %

\definecolor{SPEQr}{rgb}{0.57,0.05,0.1} %

\def\sec#1{\section{\textcolor{SECOL}{#1}}}
\def\ssec#1{\subsection{\textcolor{SSECOL}{#1}}}
\def\sssec#1{\subsubsection{\textcolor{SSSECOL}{#1}}}

\def\bay{\begin{array}}
\def\eay{\end{array}}
\def\bit{\begin{itemize}}
\def\beit{\begin{itemize}}
\def\eit{\end{itemize}}

\def\ln{\text{ln} }

\def\floor{\text{floor} }

\def\dd{~ \rotatebox{320}{\hspace{-5pt}\vbox to 5 pt {\hspace{-5pt} \hbox to 5pt {$\cdots$}}}\!\! }

\graphicspath{{}}

\definecolor{darkgreen}{rgb}{0.0,0.66,0.0} 
\definecolor{darkred}{rgb}{0.80,0.0,0.0} 
\definecolor{darkblue}{rgb}{0.00,0.0,0.6} 
\definecolor{brown}{rgb}{0.7,0.5,0.2}

\newcommand\READY[1]{\textcolor{black}{#1}}

\begin{document}

\sf 
\coloredtitle{Investigations of the exchange energy of neutral atoms in the large-$Z$ limit}
\author{\color{CITECOL} Jeremy J. Redd}
\affiliation{Department of Physics, Utah Valley University, Orem, UT 84058, USA}
\author{\color{CITECOL} Antonio C. Cancio}
\affiliation{Department of Physics and Astronomy, 
Ball State University, Muncie, IN 47306,  USA; accancio@bsu.edu}
\author{\color{CITECOL} Nathan Argaman}
\affiliation{Department of Physics, Ben-Gurion University, Beer-Sheva 84105, Israel;\ Department of Physics, Nuclear Research Center---Negev, P.O. Box 9001, Be'er Sheva 84190, Israel}
\author{\color{CITECOL} Kieron Burke}
\affiliation{Departments of Physics and Astronomy and of Chemistry, 
University of California, Irvine, CA 92697,  USA}
\date{\today}
\begin{abstract}
The non-relativistic large-$Z$ expansion of the exchange energy of neutral atoms provides an important input to modern non-empirical density functional approximations.  Recent works report results of fitting the terms beyond the dominant term, given by the local density approximation (LDA), leading to an anomalous $Z\ln{Z}$ term that can not be predicted from naive scaling arguments.  Here, we provide much more detailed data analysis of the mostly smooth asymptotic trend describing the difference between exact and LDA exchange energy, the nature of oscillations across rows of the periodic table, and the behavior of the LDA contribution itself.  Special emphasis is given to the successes and difficulties in reproducing the exchange energy and its asymptotics with existing density functional approximations.  
\end{abstract}

\maketitle
\def\floor#1{{\lfloor}#1{\rfloor}}
\def\sm#1{{\langle}#1{\rangle}}
\def\dis{_{disc}}
\newcommand{\Z}{\mathbb{Z}}
\newcommand{\R}{\mathbb{R}}
\def\w{^{(0)}}
\def\w{^{\rm WKB}}
\def\II{^{\rm II}}
\def\hd#1{\noindent{\bf\textcolor{red} {#1:}}}
\def\hb#1{\noindent{\bf\textcolor{blue} {#1:}}}
\def\eps{\epsilon}
\def\ew{\epsilon\w}
\def\ej{\epsilon_j}
\def\upet{^{(\eta)}}
\def\ejeta{\ej\upet}
\def\tjeta{\tj\upet}
\def\bej{{\bar \epsilon}_j}
\def\ewj{\epsilon\w_j}
\def\tj{t_j}
\def\vj{v_j}
\def\F{_{\sss F}}
\def\xt{x_{\sss T}}
\def\sc{^{\rm sc}}
\def\al{\alpha}
\def\ae{\al_e}
\def\bj{\bar j}
\def\bz{\bar\zeta}
\def\eq#1{Eq.\, (\ref{#1})}
\def\cN{{\cal N}}

\sec {Introduction} 
Almost a century of painstaking physical and mathematical work has proven that the
asymptotic expansion of the non-relativistic energy of the neutral atom
is
        \ben
        E \to  -c_0\, Z^{7/3} + Z^2/2 - c_1\, Z^{5/3} +...~~~~(Z\to\infty),
        \label{E}
        \een
where $Z$ is the nuclear charge (in Hartree atomic units)\cite{E88}.
Remarkably, the simplest density functional approximation, that of Thomas-Fermi theory\cite{T27,F28}, yields
precisely the leading term, allowing $c_0$ to
be calculated to arbitrary accuracy from the solution of the
Thomas-Fermi differential equation for neutral atoms\cite{LCPB09}.  In fact, Lieb and Simon
proved that Thomas-Fermi theory becomes relatively exact for the total energy
of any electronic system in a carefully defined semiclassical limit\cite{LS73,LS77}.

Modern electronic structure calculations are dominated by Kohn-Sham density
functional theory\cite{KS65}, in which only the exchange-correlation energy, $E\xc$,
need be
approximated as a functional of the density.  It has been conjectured (and proven
under various assumptions) that, in the same limit, $E\xc\LDA$ becomes relatively
exact\cite{S81,C83,E88,EB09,KR10,BCGP16,CCKB18,CPKK23}.  For exchange alone,
\ben
E\x \to - d_0 Z^{5/3} + ... (Z \to \infty),
\een
where $d_0=9c_1/11$ and is given exactly by the local density approximation (LDA) for exchange (given by the Dirac model\cite{D30}), applied to the TF density\cite{LCPB09}.
It has further been found numerically that several popular
generalized gradient approximations (GGAs) are quantitatively accurate
for the leading correction to LDA.\cite{EB09}
The construction of several approximate
semilocal functionals, including
PBEsol\cite{PRCV08}, SCAN\cite{SRP15}, and acGGA\cite{CCKB18}, have 
incorporated 
these insights.
Unfortunately, even the analytic form of these leading corrections is unknown.
Kunz and Rueedi~\cite{KR10} have discussed possible
higher-order terms in the asymptotic expansion
for exchange, distinguishing between a series with a smooth dependence on $Z$,
and oscillating terms which appear in higher orders.\cite{ESc85} They argue the need for smooth corrections to the leading order term in Eq.~2 of order $Z^{4/3}$ and $Z \ln{Z}$ terms but did not calculate the associated
coefficients or provide a proof for their existence.

\begin{figure}
\includegraphics[trim=0cm 0cm 0cm 0cm,clip=true,width=1.0\linewidth]
              {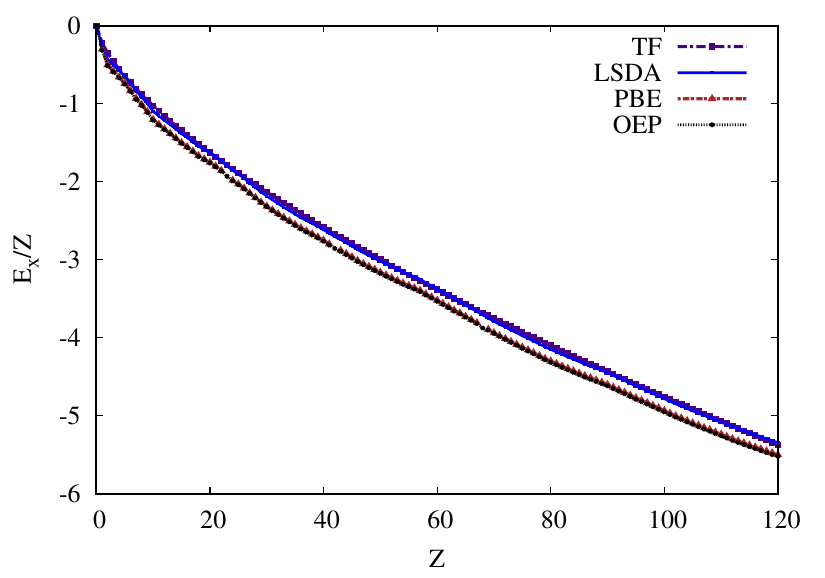}
\caption{Exchange energy per electron versus nuclear charge $Z$ for neutral atoms up to $Z=120$,
exactly (black), within LDA(blue),and with PBE (red).  The leading
order term in the large-$Z$ expansion, Dirac exchange applied to the TF
density, is shown with violet dot-dashed line.}
              \label{fig:exbasics}
       \end{figure}
Fig.~\ref{fig:exbasics} illustrates just how devilishly difficult it can be 
to find these leading corrections.   The figure shows 
exchange energy per electron as a function of $Z$.   
It includes the simple asymptotic behavior, the local density
approximation (LDA) and a popular GGA, as well as highly accurate energies
from OEP calculations.
Hidden in these barely distinguishable
lines are all the shell structure of the periodic table, and very subtle trends in the
differences.   
On this 
scale, 
the leading order term of the expansion, derivable from 
the primitive TF-Dirac model for exchange does quite well,
and all the 
complications of modern functionals only amount to a small perturbation.
This illustrates the challenge of the task of extracting accurate and
correct asymptotic behaviors, when even the form to be used is 
uncertain.

Nevertheless, there have been a number of attempts to elucidate the 
form and coefficients by careful numerical analysis of the available data.
The quantity $\Delta E\x = E\x-E\LDA\x$
is much smoother and so simpler to model than $E\x$ itself.
Original works of this sort~\cite{PCSB06,EB09} used an overly simple 
\READY{model based on a naive scaling analysis of the gradient
expansion correction to LDA that neglected changes
in the density as a function of $Z$.}
\READY{This assumes that the form of the leading correction to $\Delta E\x$ was
in simple powers of $Z^{1/3}$, neglecting the possibility of non-analytic contributions.}
In the range of $Z$ between
10 and 56 and with only noble gas atoms, this fit is indeed quite reasonable 
numerically. 

Recent work~\cite{DKGSG22,ARCB22,DKBSG23} has revealed a flaw in this 
approach.  The gradient expansion correction to the LDA
energy diverges on the TF density.  Correction of the TF density
near the nucleus
leads to a leading order term of $Z\ln{Z}$.
Based on this insight, recent work~\cite{ARCB22} revisited 
exchange data for atoms, this time up to $Z=120$, finding a remarkably 
efficient fit
of $\Delta E\x$ of closed shell atoms to the form $-Z(B\ln{Z} + C)$.
The leading coefficient is several times larger than that predicted by the
gradient expansion of exchange\cite{SB96} applied to this system.  The 
fit matches atomic exchange energies down to $Z=1$, and the coefficients
are independent of fit range and details.
An explanation for this result comes from analysis 
of the Bohr atom\cite{HL95}, consisting of noninteracting electrons in a Coulomb
potential.   
This analysis both validates the need for a $Z\ln{Z}$ term
and provides an explanation for the ratio of the numerically
observed $B$ coefficient to that of the gradient expansion.\cite{ARCB22,DKBSG23}

The older results have been used in the construction of several recent
approximate functionals.   Such fits yield accurate results for the limited
range of the fit, and hence approximations trained on such fits are accurate
for large $Z$ atoms.  But the new analysis complicates 
the picture, especially
as two coefficients are now needed to acheive the same accuracy.
The complicated behavior of the density as a function
of $Z$ makes the relationship between semilocal functionals of the density
complicated, with density functional errors harder to diagnose and fix.  It also
leads to a complicated expansion picture for LDA exchange, with many
terms beyond the $Z^{5/3}$ term characterizing the asymptotic limit, and 
oscillations as a function of the fractional filling of rows of the periodic
table.

The present work presents a detailed study of both the oscillatory $E\LDA\x(Z)$ data and the much smoother $\Delta E\x(z)$ results, with special attention given to the successes and difficulties in reproducing the exchange energy and its asymptotics with existing density functional approximations.

Organization of the rest of this paper is as follows: Sec.~\ref{sec:background} presents background
of our study, including an overview of the asymptotic expansion of exchange
of atoms and an analysis of the gradient expansion, including higher order
terms.  Sec.~\ref{sec:methods} briefly reviews numerical methods, Sec.~\ref{sec:BL} presents a comparison
of the beyond-LDA contribution $\Delta E\x$ for exact exchange and common
GGA functionals, Sec.~\ref{sec:LDA} presents our results for LDA exchange, and finally,
Sec.~\ref{sec:outcomes} presents outcomes and conclusions.

\sec{Background}
\label{sec:background}
The connection between asymptotic expansions and modern density functionals for
exchange-correlation was first discussed in Ref \cite{PCSB06}, with the appropriate
formalism illustrated for the one-dimensional kinetic energy in Ref \cite{ELCB08}.
The key analytic insight is provided by the scaling procedure of the TF
model, expressed in theorems developed by Lieb and Simon.~\cite{LS73,LS77}
The Lieb-Simon limit is approached by simultaneously
scaling the potential by a factor
$\zeta$ tending to $\infty$ and changing the particle number. 
For any potential $v(\br)$, define a $\zeta$-scaled potential
$v_\zeta(\br) = \zeta^{4/3}v(\zeta^{1/3}\br)$, 
and simultaneously replace $N$, the electron number, with $\zeta N$,
choosing $\zeta$ so the latter remains an integer.
This applies
to all atoms, molecules, and solids.  
For any finite interacting electronic system, the expansion will have the same form as
Eq. 1, but with different, system-dependent coefficients.
Because $N$ changes, this can be a challenging limit to study in practice, and
almost all numerics for interacting systems
have been extracted solely in the simple case of neutral atoms,
where $\zeta$-scaling is equivalent to changing $Z$, keeping $N=Z$.
The Lieb-Simon theorem states that, for any electronic system, TF theory
yields the leading order term (the $Z^{7/3}$ contribution) exactly.

For atoms, Schwinger first showed\cite{S81} in convincing detail that LDA exchange
yields precisely the dominant term, with many further details extracted
with Englert\cite{ES82,ES85}.  
 Later, Conlon~\cite{C83}, followed up with greater mathematical rigor by Fefferman
and Seco\cite{FSb94} gave a general proof for arbitrary systems.

Beyond the leading order term, however, little is known for certain, even
the form of the large-$Z$ expansion for the exchange energy.  Kunz and 
Rueedi\cite{KR10} found success for 2D quantum dots considered as artificial
atoms, but crucially, these lack the complicating factor of the singular
Coulomb potential.
We take their conjecture that 
the large-$Z$ expansion of the exchange energy of atoms 
has at its basis a form that is a smooth function of $Z$:
\ben
      E\x(Z)\to  -\left( d_1 Z^{5/3} + A\x Z^{4/3} + Z(B\x \ln{Z} + C\x)\right)+ \cdots
\label{ExZ}
\een
as $Z\to\infty$.
\READY{However these coefficients are expected to vary with each column of the
periodic table, leading to oscillations in energy across each row of the 
table.}  
Thus, for an accurate description of the exchange energy of every atom, the smooth form must 
be augmented by an oscillatory piece, expressed (at least in part)
in terms of coefficients that depend
on the fraction of filled shells.\cite{ESc85,E88}  
\READY{
This is straightforward in the case of the kinetic energy of Bohr atoms\cite{E88,BCGP16}, but is much more complex for the real periodic table.}
Fortunately, as we show below, and
as discussed in Refs.~\onlinecite{EB09, DKGSG22, ARCB22}, 
these oscillations are to a large degree not 
relevant to density functional development. 

Now define the beyond-asymptotic exchange energy
for each $Z$ as:
\ben
E\x\BA(Z)=E\x(Z)+d_1 Z^{5/3},
\label{ExBA}
\een
so that for large $Z$, ignoring oscillatory effects,
\ben
\frac{E\x\BA(Z)}{Z} \to -\left( A\x Z^{1/3} + B\x \ln{Z} + C\x\right)+ \cdots.
\label{ExZBA}
\een
\begin{figure}
\includegraphics[trim=0cm 0cm 0cm 0cm,clip=true,width=1.0\linewidth]
    {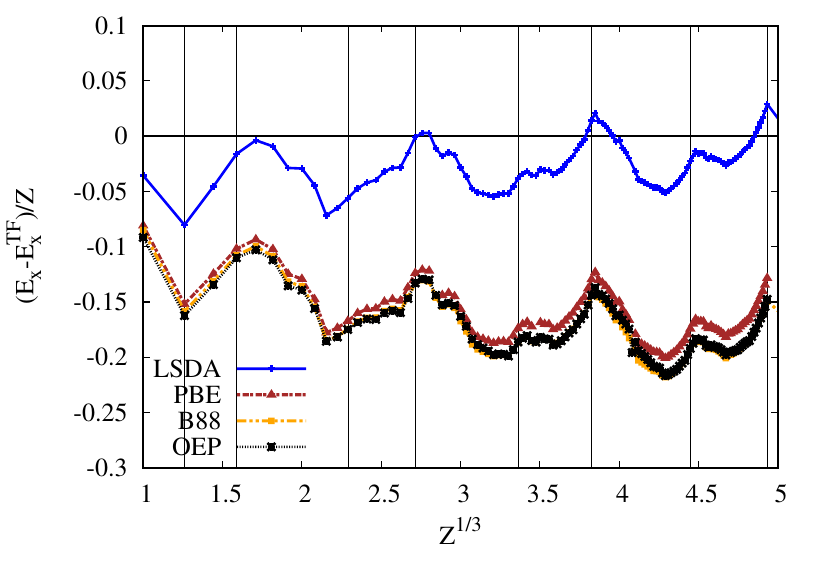}
\caption{Asymptotic residual of exchange.  Shows what is left over of the
exchange energy/particle when the leading term in the asymptotic expansion
for exchange $-(9/11)0.2699Z^{2/3}$ is removed.  Shown is the OEP data 
from opmks, the LDA, B88 exchange, PBE, and an asymptotically corrected PBE
from Ref.~\onlinecite{BCGP16} that might be deleted. The line at zero gives the Thomas-Fermi
limit of the exchange energy.  The vertical lines shows the location
of each atom with a filled $s^2$ valence shell (He and the alkali earths.)
}
\label{fig:exasy}
\end{figure}
Figure 2 shows $E\x\BA(Z)$ exactly and for various approximations.  Already,
the simple subtraction of the leading asymptotic contribution highlights
the differences and makes the periodic structure visible.   It also illustrates
a relatively slow average variation with $Z^{1/3}$.

LDA has oscillations strongly correlated with those seen in
exact exchange but a significantly different smooth contribution.
Removing this contribution from the asymptotic expansion removes
much of the complications of this periodic structure
while retaining the target of beyond-LDA density
functional models of exchange.
Thus, assuming the same qualitative behavior, we define the 
local beyond-asymptotic (LBA) energy as
\ben
   E\x\LBA(Z) = E\x\LDA(Z) + d_1 Z^{5/3},
\een
with a smooth large-$Z$ expansion of
\ben
\frac{E\x\LBA(Z)}{Z} \to  -\left( A\LDA\x Z^{1/3} + B\LDA\x \ln{Z} + C\LDA\x\right)+ \cdots,
\label{ExZ}
\een
as LDA surely also yields the correct leading asymptotic term.
Their difference $\Delta E\x(Z) = E\x(Z)-E\x\LBA(Z)$, 
the beyond-local contribution to exchange,
has an expansion
\ben
      \frac{\Delta E\x (Z)}{Z} \to  -\left(\Delta A\x Z^{1/3} + \Delta B\x \ln{Z} + \Delta C\x\right)+ \cdots.
\label{dExZ}
\een
for large $Z$.  Here, then, $\Delta A\x = A\x - A\x\LDA$, etc.
       \begin{figure}
              \includegraphics[trim=0cm 0cm 0cm 0cm,clip=true,width=1.0\linewidth]
              {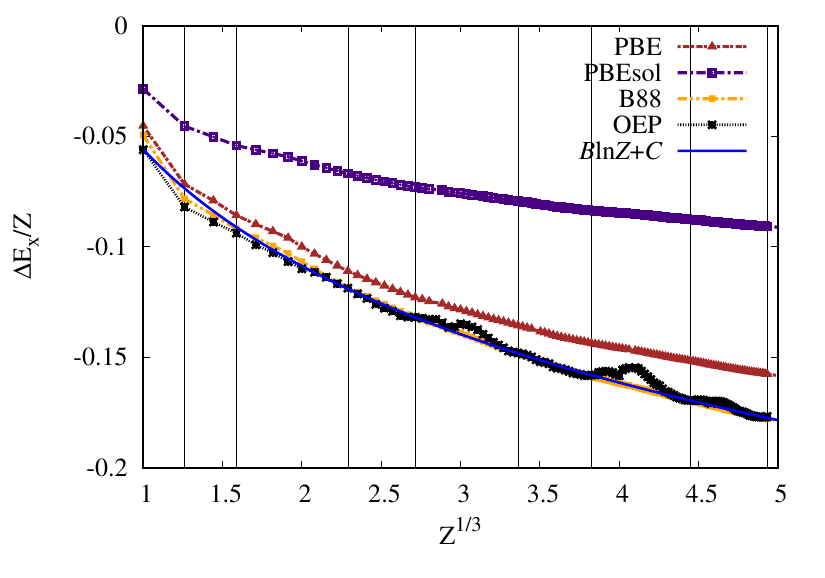}
              \caption{Difference with LDA or beyond-LDA exchange energy per electron.
                     The details are the same as the previous two figures. 
              }
              \label{fig:postLDA}
       \end{figure}
Figure 3 shows the difference between LDA and exact energies, and is clearly
far smoother than Fig. 2.  As plotted, $\Delta E\x/Z$ has a 
form $\Delta A Z^{1/3} + \Delta B\ln{Z} + \Delta C$ that lends itself to visual analysis,
and it is apparent that the trend is much more likely that of a log curve, than, 
for example, a straigbt line.
As noted by Elliot and Burke in their 
original work\cite{EB09}, the approximate
exchange energies of PBE\cite{PBE96} and of Becke 88 (B88)\cite{B88}, 
the exchange component of BLYP, 
follow very closely the smooth asymptotic trend of the exact data.
(Unfortunately, they plotted data versus $Z^{-1/3}$, and not $Z^{1/3}$, a
a strategy optimized for extracting asymptotic coefficients, but obscuring
the trend in the asymptotic form.)
\READY{A third density functional, PBEsol,\cite{PRCV08} is also shown as 
a proxy for the second-order gradient expansion which it approaches
for a system with a slowly varying density.
The plot shows clearly the underestimate of the gradient expansion
of the beyond-LDA exchange as compared to OEP, and the much
better reproduction of this data achieved by PBE and B88 by breaking 
this constraint.}

\begin{figure}
\includegraphics[trim=0cm 0cm 0cm 0cm,clip=true,width=1.0\linewidth]
                {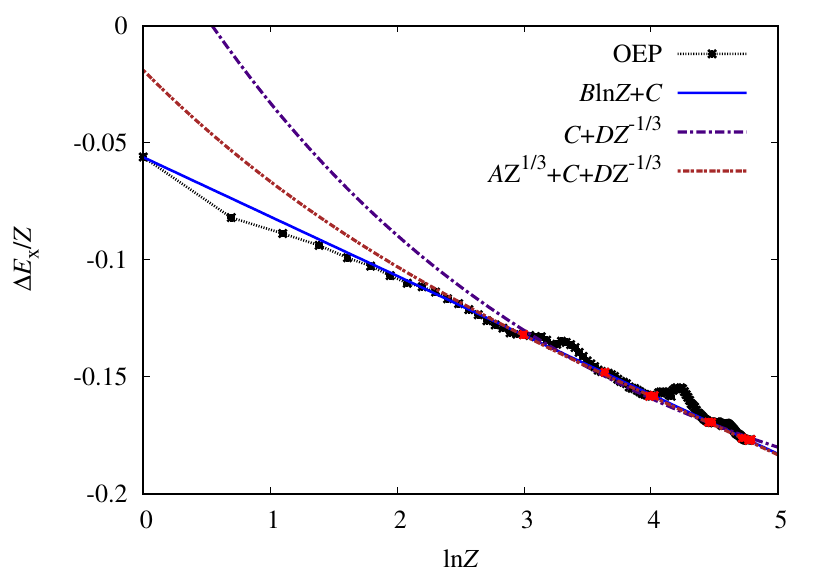}
\caption{Beyond-LDA exchange energies per electron. 
    OEP is optimized effective potential, 
   the other three curves are fits to various 
   asymptotic models as described in the text.  OEP data used to make the 
   fits are highlighted in red.
}
              \label{fig:fitlogz}
\end{figure}

Fig.~\ref{fig:fitlogz} confirms the logarithmic behavior of the 
leading order term of $\Delta E\x/Z$.
It shows the results of fits assuming leading order terms
in $\Delta E\x/Z$ to be $Z^{1/3}$ (brown), $\ln{Z}$~(blue) or constant~(violet), 
or leading order coefficients of $\Delta A\x$, $\Delta B\x$ or $\Delta C\x$ respectively, plotted
versus $\ln{Z}$.
Each fit is on a set of closed-shell atoms highlighted in red on the figure;
they are described in Ref.~\onlinecite{ARCB22} and
partly shown in Fig.~1 of the same.  
The fits are
all reasonably close to data within the range of $Z$ values, $\ln{Z} \geq 3$, of the fit
set.
However, only the logarithmic model has predictive power 
outside this range -- in fact extrapolating almost exactly down to $Z=1$.
It is not surprising that varying the fit
set of atoms (including using all the data) yields no statistically significant
change in this fit.  
We are thus confident that an expression for $\Delta E\x$ of the form
\ben
  \Delta E\x^{fit} \to -(\Delta B\x Z\ln{Z} + \Delta C\x Z)
  \label{eq:delexasy}
\een
gives an accurate description of exact exchange for the data available.

\ssec{Application to KS DFT}
\label{sec:KohnSham}

In KS DFT, we are required to give $E\x$ as a functional of the density.
As $\zeta$ becomes large, 
the exchange functional approaches the local density approximation:
\ben
    E\x[n_\zeta] \to \int d^3r\ e\x\LDA(\n_\zeta(\br)) =  - c\x \int d^3r\ n_\zeta(\br)^{4/3},
\label{ExLDA}
\een
where $c\x = (3/4)(3/\pi)^{1/3}$.  As the density also weakly approaches
the TF density, evaluation on that density
yields the coefficient $d_1$ in Eq.~2
\cite{LCPB09}.
Next, we equate $\Delta E\x(Z)$ defined in
Eq.~(\ref{dExZ}) with the difference between the LDA and exact exchange
functional
\ben
\Delta E\x(Z)=E\x[\n_Z]-E\x\LDA[\n\LDA_Z],
\een
where $\n_Z$ is the exact KS density for the $Z$ electron system
and $\n\LDA_Z$ the LDA density for the same.  This formulation may be
extended to approximate exchange functionals.
Almost all those in modern use (including all those discussed here)
recover LDA in the uniform (or large-$Z$) limit, so that $\Delta E\x/E\x \to 0$ as 
$s^2 \to 0$.

\sssec{Gradient expansion}

The natural next step beyond the local density approximation is the gradient expansion.   For a slowly-varying (infinitely extended) electron gas, this is the expansion of its energy in 
increasing (higher-order) gradients of the density.
This expansion is well-defined, and is likely asymptotic when applied to any
gas of slow but finite variation\cite{DG90}. 
The application of this expansion to finite systems as
an approximate density functional is called the gradient expansion approximation, GEA.  

For atoms, most of the density becomes slowly varying in the LS limit, i.e.,
the local dimensionless gradient vanishes as the limit is approached.  
But the region near the nucleus and the vicinity of the evanescent region,
do not. As the nucleus is approached, the GEA produces
an anomalous contribution to kinetic energy of order
$Z^2$ (but not the correct -1/2 coefficient)\cite{LCPB09}
For exchange, as we show below, GEA produces an anomalous $Z\ln{Z}$ term, 
suggesting  that such a term might be present in the exact functional.  
Moreover, modern semilocal approximations can be expanded
in the slowly-varying limit, although many do not recover the correct coefficient
for the slowly-varying gas.  Thus both the exact and most approximate functionals 
can be expected to manifest this logarithmic behavior for exchange for large $Z$.

For exchange, we write the gradient expansion
in terms of second-order $\Delta E\x^{(2)}$ and
fourth order $\Delta E\x^{(4)}$ terms as
    \ben
        E\x\GE = E\x\LDA + \Delta E\x^{(2)} + \Delta E\x^{(4)} + \cdots.
    \een
The second-order gradient correction is of most importance here and is given by 
    \ben
        \Delta E\x^{(2)} = \mu_2 \int s^2 e\x\LDA(\n(\br)) d^3r
        \label{eq:gea2}
    \een
with $\mu_2 = 10/81$~\cite{SB96} and $s^2$ is a scale-invariant gradient given by
            \ben
                s^2 = \frac{|\nabla n|^2}{4 k_F^2 n^2}
            \een
where $k\F = (3\pi^2n)^{1/3}$ is the local Fermi wavevector.
To generate the leading beyond-LDA term in the exchange expansion, 
one can first try a naive scaling argument.  
Apply the second-order gradient expansion to the TF density $n\TF(\br)$.  Note~\cite{LCPB09}
that for any finite value of $r$,
$s^2 \sim Z^{-2/3}$ for the TF density, while the LDA gives a factor
of $Z^{5/3}$.  The end result
of Eq.~\ref{eq:gea2} should give 
the order of the first term beyond leading order, and is of order $Z$.

To obtain a logarithmic term from the gradient requires more careful analysis 
of the contribution from the very high density inner core of the large-$Z$ atom.\cite{ARCB22}
Here, $n\TF(r)$ diverges as $1/x^3$ and $s^2$ diverges as $1/x$, where 
$x = Z^{1/3}r/a$ is the scaled distance derived from TF theory\cite{E88}, 
and $a = (1/2)(3\pi/4)^{2/3}$.  
Thomas-Fermi theory breaks down for $s^2 \sim 1$ and care must be taken to 
ensure finite integrals.  Including the divergence of the TF density
at the nucleus produces divergent integrals.  For the region inside
$x \sim 1$, the location of the peak of the radial density 
probability, we have
            \bea
               n(r) & \sim & \frac{Z^2}{4\pi a^3} x^{-3/2} \\
               e\x\LDA d^3r & \sim & Z^{5/3} \frac{c\x}{(4\pi)^{1/3}} dx \\
               s^2 & \sim & \frac{a_1^2} { Z^{2/3}} x^{-1} 
            \eea
where $a_1 = (1/2)(9/2\pi)^{2/3}$~\cite{LCPB09}.
With the resulting integrand diverging as $1/x$ as $x \to 0$,
$\Delta E\x^{(2)}$ diverges logarithmically.  
The natural small-radius cutoff is $r \sim 1/Z$, the distance from the 
nucleus in units of $a_B$ inside which TF theory fails.
This translates to $x \sim Z ^{-2/3}$ so we find a net contribution to the 
leading order of
    \ben
    \Delta E\x^{(2)}  \sim \int_{Z^{-2/3}}^1 x^{-1} dx = \frac{2}{3} Z\ln{Z}
      + \mathcal{O}(Z).
    \een
Restoring constants produces 
           \ben
                \Delta E\x^{(2)} \sim \Delta B\GEA_2 Z\ln{Z} + \Delta C\GEA_2 Z,~~~~~ Z \to\infty
                \label{eq:ourmodel}
           \een
with
           \ben
               \Delta B\GEA = \frac{3}{4\pi^2}\mu_2.
               \label{eq:deltabgea}
           \een

To see if higher orders in the gradient expansion alter
this coefficient, we proceed with the
fourth-order contribution in a similar fashion.  
The general form is
    \ben
        \Delta E\x^{(4)} = \int ( \mu_{pp} p^2 + \mu_{pq} pq + \mu_{qq} q^2) e\x\LDA  d^3r,
    \een
where the $\mu$'s are known coefficients.
Here, $p=s^2$ 
and $q=\nabla^2 n/ 4k_F^2 n$ is the scale-invariant 
Laplacian of the density.
Since $q = p/3$
for the region of space where the logarithmic divergence in
$\Delta E\x^{(2)}$ occurs,\cite{LCPB09} we find
           \ben
             \Delta E\x^{(4)} \sim \mu_4\, \int d^3r p^2 e\x\LDA,
           \een
where $\mu_4=\mu_{pp}+\mu_{qp}/3+\mu_{qq}/9$.
An analysis similar to that for 2nd-order
 shows that the integral scales naively 
as $Z^{1/3}$, but the integrand varies as $1/x^2$.
The overall behavior is
         \ben
             \Delta E\x^{(4)} \sim Z^{1/3} \int_{Z^{-2/3}x_1}^{x_2} x^{-2}dx
                 \sim Z/x_1
              \label{eq:deltaex4}
         \een
In this case we can't eliminate a constant $x_1$ that determines
the exact small-radius cutoff, but a good criterion for this 
can be found, as we show below.   More important is to note that
neither the fourth-order GE, nor any high-order
order term in the gradient expansion contributes to the logarithmic
contribution $\Delta B\x$ to $\Delta E\x$: it only comes only from the 
second-order gradient expansion as far as we can tell.

Thus, the minimal density functional approximation
that captures the leading order term in exchange
beyond the LDA is the GGA, which has the general formulation given by
       \ben
       E\x\GGA = \int d^3r F\x(s^2) e\x\LDA,
       \een
which typically reduces to the form of the
second-order gradient expansion, $F\x \sim 1 + \mu s^2$,
in the slowly-varying limit, $s^2 \to 0$.
As we show in Sec.~\ref{sec:meta-GGAs},
meta-GGAs only add
significant corrections to this form to fourth-order and are less
relevant here.
Different GGA's produce different $\Delta B\x$ coefficients because of differing
values for $\mu$.  
Thus the PBEsol~\cite{PRCV08} uses the gradient expansion coefficient,
$\mu = 10/81\sim 0.123$, but PBE~\cite{PBE96} has
$\mu = \beta \pi^2 /3 \sim 0.21951$ based on a different choice of constraints, 
and B88~\cite{B88}, 
has $\mu = 0.275$.  
But our asymptotic model [Eq.~(\ref{eq:ourmodel})] of actual atom 
exchange energies 
has $\mu \sim 1/3$, considerably larger than all these.

This discrepancy indicates the limitation (first pointed out in Ref.~\onlinecite{PCSB06})
of the gradient expansion as a model for Coulombic systems.  The 
gradient expansion is derived  for slowly-varying gases, without classical turning
points at the Fermi level -- something
that is not true for any atom with finite $Z$.  In addition to the Coulomb singularity that
ensures there will be a finite region near the nucleus where Thomas-Fermi analysis fails,
the classical turning point at the valence edge necessarily requires corrections
of fundamentally different form than the gradient expansion, as discussed in Ref. \cite{OB21}.
Indeed, the Scott term in the total energy\cite{S52}, of order $Z^2$ [Eq.~(\ref{E})], 
is evidence of such
corrections. More precisely, a term $\sim Z^2$ is generated from the gradient expansion 
for the total energy by using an analysis of the inner core equivalent to that discussed here
for exchange, but the coefficient has the wrong sign and magnitude.  The correct coefficient
of -1/2 is easily deduced by the direct analysis of the Bohr atom however.\cite{BCGP16}. 
It seems likely, therefore, that the same thing is happening for exchange: the gradient expansion
indicates the need for an anomalous $Z\ln{Z}$ term but does not determine the correct coefficient.

Based on this insight, Ref.~\onlinecite{ARCB22} applied an analysis of the 
exchange energy of the Bohr atom to find that the exact exchange indeed had a 
logarithmic $Z\ln{Z}$ term, with a coefficient that was exactly 2.7 times the value 
obtained by the GEA applied to that system.  Assuming that the same ratio should hold 
for real atoms, yields a correction of $2.7\Delta B\GEA$ [Eq.~(\ref{eq:deltabgea})] or 
$1/4\pi^2$.
Further requiring, with probably somewhat less reasonability, that  
$Z=1$ should yield the exact $E\x$ for hydrogen produces the following conjecture
for the beyond-local contribution:
            `   \ben
   \Delta E\x  = -Z\left(\ln{Z}/(4\pi^2) + 5/16 - 0.2564\right).
                   \label{eq:delextheory}
                \een
This remarkably matches our numerical parameters within their statistical error. 

        \sec{Methods and numerical checks}
           \label{sec:methods}

In the following we must
distinguish between trends with $Z$, $Z\log Z$ and
$Z^{4/3}$.  To do so, we extend our data set to as high a $Z$ as possible,
ignoring issues of experimental stability and relativity, a task that 
involves two different atomic DFT codes.
First we use the optimized effective potential code opmks~\cite{ED99} to 
perform nonrelativistic exact exchange \READY{and spin-density dependent,
self-consistent density functional}
calculations for all neutral atoms up to $Z=120$.  
Unfortunately, the inversion problem used to find the potential in the OEP
begins to fail for $Z>120$ and  DFT calculations using OPMKS fail to 
converge reliably for $Z>362$.
         
For GGAs and the LSDA for large-$Z$ atoms ($Z > 120$) we use the atomic 
pseudopotential code FHI98PP\cite{FHI98PP} in its all-electron, 
non-relativistic mode. 
\READY{
This code enables us to make GGA and LDA calculations
extending the periodic table to $Z=978$ without significant 
sign of numerical stress. 
This corresponds to 16 full shells of the periodic table, and one 
filled 17s shell -- the alkali earth metal with valence shell 17s$^2$.
(A few atoms had frontier orbitals
that failed to converge properly, mostly open-shell atoms and
two closed-shell atoms of the extended periodic table with frontier
orbitals with a high angular momentum.)
}
         
FHI98PP computes wavefunctions on a grid of radial points, with spacing 
between successive points increasing by a geometric factor $\gamma$.
We use the default grid which starts at $r=0.00625/Z$, 
with a step size that increments by a growth factor $\gamma=1.0247$ out to a 
maximum radius of 80. 
To check the quality of FHI98PP energies, the highest occupied orbital 
energy eigenvalues were compared to those of OPMKS
for closed shell atoms with $Z<362$ using LSDA exchange.  These calculations were done both using a finer and more course grid than the default.
The results are indistinguishable within machine precision.
The details of the stress test can be seen in the supplementary materials.

The shell structure of the large-$Z$ atoms was assumed to follow the Madelung principle for closed shell atoms.  The validity of this extension has been tested by comparing the total energy of Aufbau-constructed shells versus several other shell configurations for 
elements 976 (filled 16p), 970 (filled 15d), and 816 (filled 16s).
For all cases tested, the Aufbau construction yields the nonrelativistic 
ground state for these atoms.     

An important potential source of error is the use of OEP to obtain 
exact exchange energies -- the true ground state has correlation energy,
which will slightly change the density and orbitals used to evaluate exchange.
We do not have a way to assess this issue directly. However, as a proxy, we
have calculated beyond-local exchange energies using the difference between
OEP and LDAX (i.e. using LDA exchange only), on the principle that this 
should cancel some of the error of not treating correlation in the OEP.  
Refitting the coefficient $\Delta B\x$ for the leading beyond-local
term in exchange leads to a value of $0.0248$, a 2\% change in the
fitted value of $0.0254$ reported in Ref.~\onlinecite{ARCB22}.  
\READY{A similar calculation, comparing OEP with PBE correlation to LSDA, yields $0.0250$.}
These results
are an order of magnitude smaller an effect than the discrepancies 
we see between the predictions for this quantity by GGA and our model, 
and is nearly within statistical error.  
       
Statistical fits used gnuplot plotting and Levenberg-Marquardt nonlinear
regression.  A list of approximate and exact
 exchange energies used in this paper is included in the supplementary materials.

\sec {Results with approximate functionals}
\label{sec:BL}

\begin{figure}
       \includegraphics[trim=0cm 0cm 0cm 0cm,clip=true,width=1.0\linewidth]
        {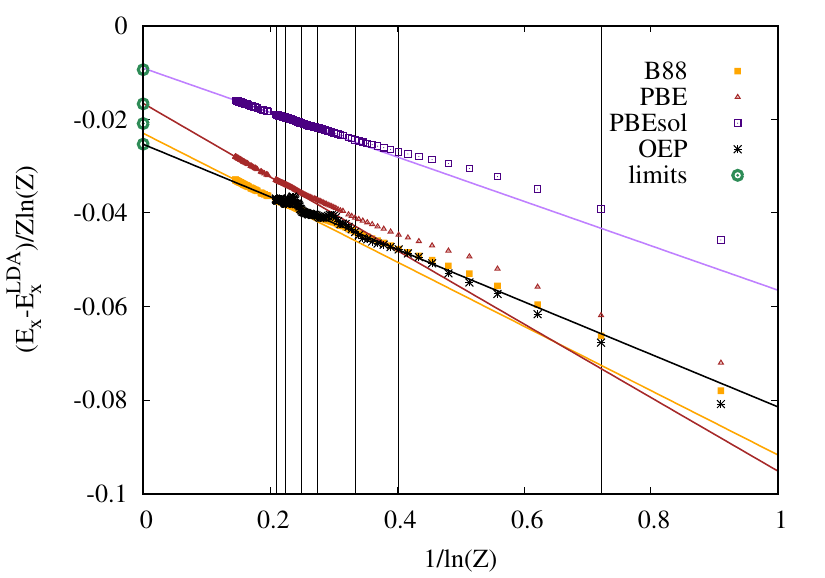}
\caption{Extrapolation of beyond-LDA exchange energies per electron to the 
$Z\to\infty$ limit for OEP data and several common GGA's.  
The $y$-axis intercept yields the coefficient $\Delta B\x$.  
PBEsol is included as a proxy for the gradient expansion.  
Black line is the semi-theoretical result of Ref.~\cite{ARCB22} while the 
other straight lines are fit to DFT models for $Z>259$.
Green dots are calculated theoretical limits of each model and the OEP.
}
          \label{fig:fitinvlogz}
\end{figure}

\ssec{Generalized gradient approximations}
How do standard DFT approximations perform  in reproducing the asymptotic behavior 
of exact exchange data?  To find out, we recast our data 
so as to extract efficiently the asymptotic character of the exchange energy.
Starting with the asymptotic form $\Delta E\x^{asy}$ [Eq.~(\ref{eq:delexasy})] 
describing the nonoscillatory contribution to the beyond-LDA exchange 
energy in the large $Z$ limit, we reframe it as 
    \ben
    \frac{\Delta E\x^{asy}}{Z\ln{Z}} = \Delta B\x + \Delta C\x x, 
     \label{eq:deltaexasy}
    \een
where $x = 1/\ln{Z}$.
Plotting $\Delta E\x^{asy}/{Z\ln{Z}}$ versus $x$ casts this 
relation as a straight line with $y$-intercept $\Delta B\x$ and slope $\Delta C\x$.
This plotting convention
yields an easy visual comparison to the behavior of exact exchange and 
of approximate functionals.  
Fig.~\ref{fig:fitinvlogz} shows the results.  
Black crosses show the beyond-LDA exchange energy for OEP, and the black line extrapolating 
to $x=0$ is the asymptotic
model Eq.~\ref{eq:delextheory}, with the $y$-axis intercept shown as a green circle.  
We place vertical lines 
at the location on the $x$ axis of the alkali
earth atoms, from 8s$^2$ ($Z=120$ or $x=0.21$) to 2s$^2$ ($Z=4$, $x=0.72$).  
Helium and hydrogen have $x>1$ and are not shown.
              
Similar plots are included on the figure 
for three different GGA's: PBEsol, PBE, and BLYP, each evaluated
on an extended data set including closed shell atoms up to $Z=978$.
The asymptotic trend to large $Z$ for each is estimated by taking a linear fit
for $x<0.18$ ($Z\geq 260$) and extrapolating to $x=0$ and shown as a straight
line.
As discussed in Sec.~\ref{sec:methods}, the theoretical large-$Z$ limit of a 
GGA is determined by the 
its coefficient $\mu$,
yielding a prediction
for $B$ [Eq.~(\ref{eq:deltabgea})] proportional to $\mu$.  The prediction
for each GGA is shown as an additional green circle on the $y$-axis.

We do not compare directly to the gradient expansion because of the large errors it suffers 
in the exponential tail of small atoms (where $s^2$ diverges to infinity). 
Instead we use PBEsol as a surrogate,
as it yields the exact second-order gradient expansion ($\mu = 10/81$)
for the slowly-varying gas.
PBEsol mimics the general trend of the OEP 
correctly, i.e., it has a reasonably close slope or $\Delta C\x$ coefficient.
But its intercept is nearly one-third too small, consistent with the finding 
in Sec.~\ref{sec:background}.  This leads to a significant error in exchange energies of atoms.

At the same time, by roughly doubling $\mu$ compared to the gradient expansion,
PBE and BLYP correct for the PBEsol errors for $Z$ of chemical 
relevance but still fall short of reaching the predicted $\Delta B\x$ from
asymptotic analyis. 
The extrapolated large-$Z$ behavior of each density functional, with the notable
exception of BLYP, is close to that obtained from the coefficient
of the second-order gradient expansion of each model.  This corroborates the
assessment in Sec.~\ref{sec:background} that the value of $\Delta B\x$
obtained in a gradient expansion is determined from the second-order 
contribution only, [Eq.~(\ref{eq:deltabgea})],  and not from higher-orders.
As we shall see below, even B88, the exchange functional of BLYP, 
can be made to comply with a little extra work.

\ssec{Asymptotic analysis of remainder}

\begin{figure}
       \includegraphics[trim=0cm 0cm 0cm 0cm,clip=true,width=1.0\linewidth]
       {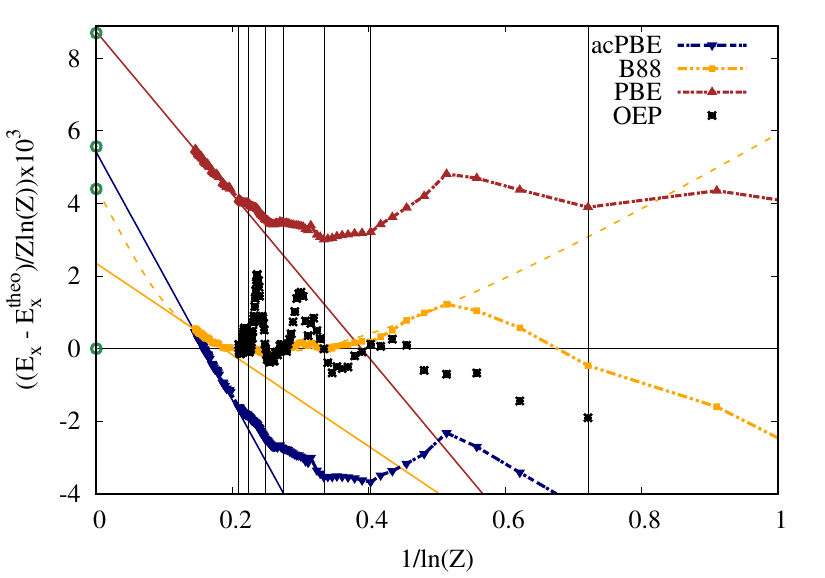}
       \caption{Extrapolation of difference between DFT exchange 
               energies-per-particle and our smooth asymptotic model to the
        $Z\to\infty$ limit.  
        acPBE has a gradient expansion
        coefficient of $\mu=0.2605$ as suggested by 
        Ref.~\onlinecite{EB09}.  Theoretical values for extrapolated
        $\Delta B\x$ coefficients shown as green dots and fits to functionals as 
        straight black lines. 
        Location of alkali earths indicated by vertical black lines as 
         guide to the eye.
       }

       \label{fig:invfit}
\end{figure}

We performed one further piece of analysis, to plot the difference 
between the exchange energy-per-particle of several approximations
and the ``nearly" theoretical model of 
Eq.~(\ref{eq:delextheory}).
This is shown for the OEP and several GGA approximations to the
exchange energy  in Fig.~\ref{fig:invfit} using the same convention for
$x$ and $y$ axes as Fig.~\ref{fig:fitinvlogz}.

        At the fine energy scale revealed by subtracting off the leading
        order terms to the beyond-LDA expansion, 
        the small oscillations in the OEP data becomes significant.  

        The figure shows that our theoretical model for the asymptotic 
        limit has, to some extent, captured the
       trend of the lower energy edge of the oscillations in OEP.  
       The upper peaks occur at filled 2p, 3d, 4f valence shells (Ne, Zn, Yb), 
       and trend to a rather different asymptote.  
       (To identify Ne, note that the bars show filled alkali earths, 
       starting with the 8s$^2$ filled valence shell and working back to 2$s^2$.  
       Ne is the peak two atoms from Mg, the vertical line at $x \sim 0.4$).  
        The amplitude of oscillation is roughly 0.003~$Z\ln{Z}$ 
        ($3Z\ln{Z}$ in mHa) in the 
        range of $Z$ values for which we have data, i.e.  an order of 
        magnitude lower than the $Z\ln{Z}$ coefficient for the smooth
        asymptotic trend.
        It is also quite possible that these oscillations grow as $Z^{4/3}$,
        as do those of the total energy~\cite{E88} and, as we shall show
        presently, those of the LDA exchange energy of real atoms.
        In comparison, the GGA's greatly underestimate the magnitude of these oscillations, although it seems that they do have small ``kinks" at roughly the
same $Z$ values as the minima in the OEP data trend.

        \ssec{Generalized gradient approximations}

We studied two GGA's with the PBE form.  PBEsol has $\mu$ matching that
of the gradient expansion, while PBE is about double that. 
        The
        impact seen in Fig.~\ref{fig:fitinvlogz} is to improve exchange energies overall,
        lowering $\Delta B\x$ and increasing the slope $\Delta C\x$.  
        What happens if we 
        should continue the trend?  Fig.~\ref{fig:invfit} 
        includes an PBE model that was asymptotic adjusted using the old
fit.   It has $\mu = 0.2609$, closer to what
        B88 exchange tends to, and suggested by Ref.~\onlinecite{EB09},
        but otherwise keeping the same functional form as the PBE and PBEsol.
        The result (downward triangles) is clearly to overshoot the 
        OEP, with much too steep a slope in the data.  Arguably, PBE already had the
        close to optimal slope in $Z$ for $10<Z<120$; 
        attempting to shift the magnitude of the correction to the OEP by 
        changing $\mu$ worsens the fit to the data as a function of 
        $\frac{1}{\log Z}$.

        Secondly, we can verify that the linear extrapolations
        to $x=0$ that we made in Fig.~\ref{fig:fitinvlogz} 
        (shown again in Fig.~\ref{fig:invfit} as solid lines) hold up, at least for the 
        PBE-like functionals.  
        Clearly, the $y$-axis intercepts nearly exactly reproduce those predicted
        from theory for the gradient expansion of each model.  This confirms
        that all these models approach their asymptotic limits fairly quickly.
        This also gives us confidence that
        we can make reasonable estimates for the slope (or the $\Delta C\x$
        coefficient in the asymptotic analysis). 
        The results are summarized in the following table.
\begin{table}
\begin{tabular}{@{}l|lrrcc@{}} 
        \hline \hline 
        Model  & $\mu$  & $\mathrm{\Delta B}\x$     & $\Delta B\x^{fit}$ & $\Delta C\x^{fit}$ & Ratio \\
        \hline 
        PBEsol & 0.1235 & 9.38  & 9.22  & 46.0 & 4.9\\
        PBE    & 0.209  & 16.68 & 16.60 & 78.5 & 4.7\\
        acPBE  & 0.2609 & 19.83 & 20.82 & 90.3 & 4.5\\
        B88    & 0.275  & 20.9  &       & 90.8 & 4.3 \\
        Theory &        & 25.33 &       & 56.0 & 2.2 \\
        \hline \hline 
\end{tabular}
\caption{
Results of analysis of the large-$Z$ limit of various GGAs described in the text.  
$\mu$ is the coefficient of the gradient expansion used in each functional, $\Delta B\x$, 
the theoretical value of the logarthimic coefficient derived from the 
gradient expansion, $\Delta B\x^{fit}$ and $\Delta C\x^{fit}$ are the 
results of fitting GGA data to Eq.~(\ref{eq:delexasy}) for PBE and its variants,
or to Eq.~({eq:delexb88}) for B88.
Ratio is the ratio $\Delta C\x^{fit}/\Delta B\x$.  Asymptotic coefficients
are reported in mHa.
}
\label{table:beyondLDA}
\end{table}
        We see that $\Delta C\x$ as well as $B$ in the PBE ``family" of GGA's varies roughly
        linearly with $\mu$, with a nearly constant ratio of $\Delta C\x / \Delta B\x$.  
        Interestingly,
        this ratio is double that predicted by our semi-theoretical model, indicating
        that no PBE-like functional could capture the basic trend in exchange data well.
        In this context, PBE is perhaps the best choice since it 
        balances errors in $\Delta B\x$ and $\Delta C\x$ roughly evenly.

We turn next to B88, which has designed on very different principles, including fitting
one parameter to the exchange energies of the noble gas series.
Clearly B88 matches OEP very well for $10<Z<120$, as is well known.  The 
main issue is a poor description of oscillations, particularly that of the 
first row atoms needed for organic chemistry.  
Thus, if we follow the protocol for obtaining $\Delta B\x$ and $\Delta C\x$ that
we used for the OEP, using the same data set of selected closed-shell atoms between
$Z=20$ and 120,
we find $\Delta B\x\B88 = 25.72(9)$~mHa, $\Delta C\x\B88 = 55.0(4)$, close to
our derived results of Eq.~(\ref{eq:delextheory}).  
(Interestingly this fit does not extrapolate as well to the B88 data in the 
first row as it does the OEP data.)

On the other hand, if we follow B88 for $x<0.2$,i.e, $Z>120$, we find it soon
begins to deviate from this trend.
Ultimately the asymptotic trend has to match a $y$ intercept consistent
with $\Delta B\x\B88 = 20.9$, the theoretical value derived by expanding the B88 
functional about small $s^2$, and roughly 4~mHa higher than the $\Delta B\x$ value
predicted by our work.  That is to say, the asymptotic trend of beyond-LDA 
exchange in B88 is to some extent only reached for $Z$'s larger than those we have
OEP data for.

To check that this is reasonable, we fit 
beyond LDA exchange for B88 to the form 
              \ben 
                 \begin{split}
                 \Delta E\x\B88 =& -Z\left( \frac{3}{4\pi^2}\mu\B88\ln{Z} + \Delta C\x\B88 \right)   \\
                    & -Z^{2/3} \left( \Delta D\x\B88 \ln{Z} + \Delta E\x\B88 \right)
                 \end{split}
                 \label{eq:delexb88}
              \een
finding $\Delta C\x\B88=90.8$~mHa, $\Delta D\x\B88= -4.18$, and $\Delta E\x\B88=-46.5$ for 
$Z>12$.  
This is shown as a yellow dashed curve and gives a plausible if imperfect 
fit to the BLYP data for $Z>12$, failing for small $Z$. The $\Delta C\x$ coefficient is 
quite plausible, yielding a $4.3:1$ ratio for $\Delta C\x/\Delta B\x$, close to that of the 
PBE-like functionals.

This analysis suggests somewhat disquietingly two possible scenarios for OEP 
exchange.  One is our picture -- that it is described by a simple and highly accurate asymptotic model one which B88 only matches within the range of data $10<Z<100$ for which 
it was fit.  Or, since our own model ultimately relies on its unusual fidelity to 
OEP data for $1<Z<120$,
the opposite might be true -- the true asymptotic behavior for 
beyond LDA OEP might not make itself apparent until $Z\gg 100$ and the 
behavior relevant for real electronic structure might involve a large
number of subdominant terms that would be quite hard to determine by
any means.  B88 or something near it could be the asymptotically
correct picture.
Ultimately, we will need a fully \textit{a priori} determination of $\Delta B\x$ to
decide between these two scenarios.

\ssec{meta-GGAs}
\label{sec:meta-GGAs}
We finish our discussion of beyond-local results with a short analysis 
of meta-GGAs.  
A meta-GGA relies on three semilocal arguments: $n(\br)$,
$\nabla n(\br)$, and $\alpha(\mathbf{r})$. 
The last of these introduces
the noninteracting KE density $\tau\s$:
           \ben
              \alpha = \frac{\tau\s - \tau\W}{\tau\TF}
           \een
where $\tau\s = \frac{1}{2} \sum_i^N \nabla \phi(\br)^2$
is the positive definite kinetic energy density, 
and $\tau\W = \frac{1}{2}|\nabla \sqrt{n}|^2$,
the von-Weizs\"acker kinetic energy density, is its form for a single orbital
system.
Among recent metaGGAs, TPSS~\cite{TPSS03} and its derivatives 
and SCAN\cite{SRP15} are
designed to recover the fourth-order gradient expansion for $E\x$ for
the slowly varying gas, while others such as r$^2$SCAN\cite{FKNP20} and MVS\cite{SPR15} do not.
But in all cases, 
as $\alpha$ modifies only the 
fourth-order gradient correction, it does not 
modify $\Delta B\x$.

There is one thing we can take from this.  
For the bulk of the atom core, $\alpha$ is closely approximated by
its gradient expansion~\cite{CR17,RC21}, and has the form~\cite{Kc57,H73}
           \ben
               \alpha \sim 1 - \frac{45}{27} s^2 + \frac{20}{9} q.
           \een
For the inner core, with scaled radius $x=Z^{1/3}r/a<1$, 
we also have $q = s^2/3$ so
$\alpha \sim 1 - \frac{25}{27} s^2$
and is a redundant parameter.
But given the constraint that $\tau\s > \tau\W$,~\cite{LO88}
we require $\alpha >0$.  
This is broken for small $x$ for the value of $s^2$ equal to
$a_1^2/(Z^{2/3}x_1) = 27/25$.
and suggests a value for the short-range cutoff parameter $x_1$ in 
Eq.~\ref{eq:deltaex4}:
             \ben
               x_1 = \frac{25}{27} a_1^2.
             \een
With this, the leading order correction due to the divergence
of $n\TF$ and $s^2$ in the inner core becomes 
             \ben
                 \Delta C\GEA_4 \sim \frac{243}{200 \pi^2} \mu_4
             \een
where
             \ben
                 \mu_4 = \mu_{pp} + \mu_{pq}/3 + \mu_{qq}/9 \sim 0.06809
             \een
using canonical values,\cite{SB96} resulting in a contribution of
about 8~mHA to $\Delta C\x$.
This is about 20\% of the overall $\Delta C\x$ obtained by numerical
analysis of atomic data.
\READY{
This reinforces our finding that the GGA itself dominates the beyond-LDA exchange energies, in that it whows that not only $\Delta B\x$ is unaffected, but also $\Delta C\x$ is affected only to a modest extent.}

\sec {Asymptotics of LDA}
\label{sec:LDA}

Surprisingly, the asymptotics of the LDA exchange energies are more difficult
to capture than the beyond-LDA corrections.  This is because LDA exchange
energies have strong oscillations across the periodic table, presumably due to
the varying nature of open shells across a row.  Because of the KS construction,
presumably LDA itself produces extremely accurate densities, with only very insignificant
changes (for our purposes) when more sophisticated approximations are used.

\begin{figure}
\includegraphics[trim=0cm 0cm 0cm 0cm,clip=true,width=1.0\linewidth]
{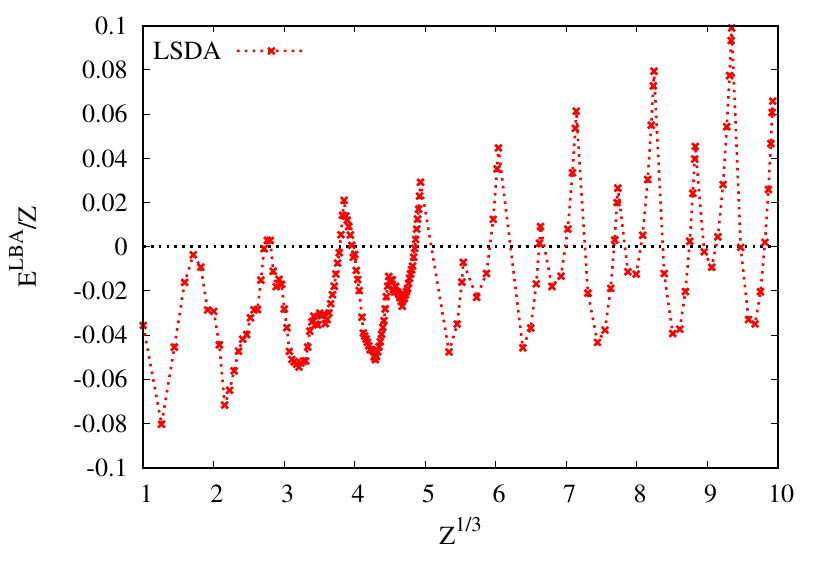}
\caption{The LSDA exchange for neutral atoms up to $Z=120$ and closed-shell neutral atoms up to $Z=978$, as described in the text, versus $Z^{1/3}$} 
   \label{fig:bigZ_LDA_no_split}
\end{figure}

       \ssec {Basics} 
              Because of the complicated shell structure seen in Fig.~\ref{fig:exasy}
              it is worthwhile to investigate the asymptotic expansion for 
              the KS LDA. Fig.~\ref{fig:bigZ_LDA_no_split} shows the difference between 
              the LSDA and asymptotic LDA energies per electron, $E\x\LBA/Z$, 
              for all atoms up to $Z = 120$ and for those
              atoms with closed s, p, d, f, g, h, i, or j valence shells up
              to $Z = 978$
              The data covers 16 rows of the extended
              periodic table and the 17s$^2$ alkali earth. 
              There is a
              clear (but complicated) oscillatory pattern, superimposed upon a
              gradual upward drift in energy.  The oscillatory pattern has 
              eight peaks indicating the period is every two rows of the periodic
              table, and the amplitude grows with $Z$.
Note that the changes within one oscillation dwarf those of the beyond
LDA term $\Delta E\x$ discussed in the previous section.
              
              Since we are considering the error in energy for a fixed functional
              form, Dirac exchange of Eq.~\ref{ExLDA}, any difference from the
              Thomas-Fermi asymptotic limit is due to density differences.  
              This figure demonstrates the effect caused by introducing 
              shell structure
              into the density, as compared to the smooth TF scaling form responsible
              for the leading order term in exchange. 
              
\begin{figure}
\includegraphics[trim=0cm 0cm 0cm 0cm,clip=true,width=1.0\linewidth]
                {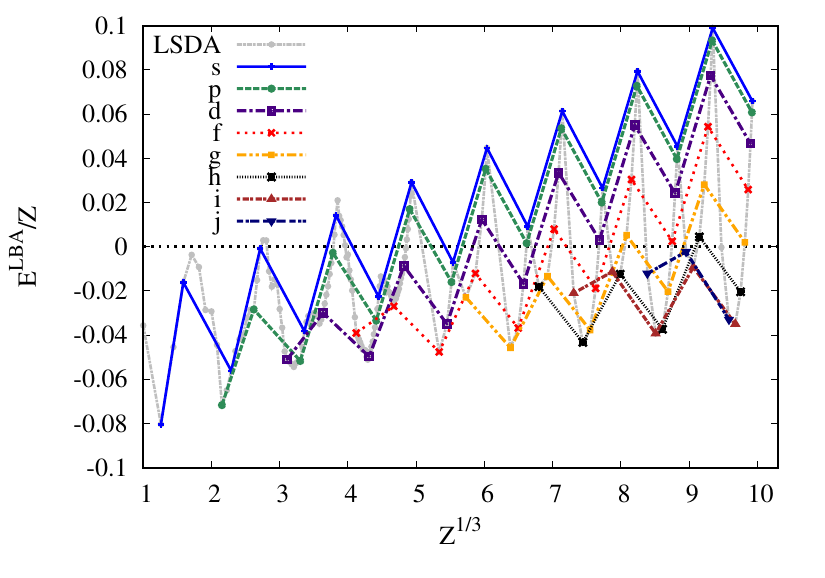}
\caption{Plots of the beyond-asymptotic LDA exchange for atoms up to $Z=120$ and all closed-shell atoms up to $Z=978$. Overlaid are energies sorted by column of the periodic table. $s$ indicates column 2
with closed $s$ valence shell (alkali earths), $p$, noble gases, $d$, 
group 12 closed $d$ shell atoms, etc. 
$g$ through $j$ refer to total angular momenta $l$ from 4 to 7, possible in the extended
periodic table.
       }
       \label{fig:bigZ_LDA}
\end{figure}

In Fig.~\ref{fig:bigZ_LDA} we recast the LDA data to bring out some of 
the hidden structure underlying the complex oscillatory patterns
exhibited by the full data set.  We do so by
highlighting
atoms that belong to columns of the periodic table 
corresponding to closed-shell systems.
These include He plus the alkali earths (with an outermost closed $s$ 
shell), noble gases (closed $p$ shells), group 12 metals (closed $d$ shells), 
and 
continuing up to closed $j$ valence-shell atoms. (The $j$ shell 
corresponds to total angular momentum $l=7$ and, according to Madelung's
rule, is first occupied in the 15th row of the extended periodic table.)  
Each column is highlighted by a line joining atoms of that column.

Each column thus displayed forms a stair case pattern that differentiates
even and odd rows of the periodic table, falling into a fairly
predictable amplitude oscillation after first one or two rows 
of each column.
\READY{
The $g$ and $h$ closed-shell series of the extended periodic table 
appear to break this pattern, but
because the numerical calculation of the first occurence of these columns
(i.e., the 5g and 6h valence shells) failed to converge.
}
The pattern resets each time a new angular momemtum is added to the
atomic configuration.

The alkali-earth staircase seems to determine approximately the 
higher energy edge of the oscillatory pattern of the atomic
exchange energy (Fig.~\ref{fig:bigZ_LDA_no_split}, and the background in 
gray in Fig.~\ref{fig:bigZ_LDA}).  
This is true both for even rows and odd, with the even rows
forming the upper limit of the overall pattern.
The low-energy edge of Fig.~\ref{fig:bigZ_LDA_no_split} is shown in 
Fig.~\ref{fig:bigZ_LDA} 
to be due to successive low-$Z$ rows of each closed shell column, the first,
an atom (He) with $s$ frontier orbitals,  then $p$ (Ne), 
then two with $d$-shell
valence (3d and 4d) and thereafter, a new value of $l\HO$ for each
minimum.
It also seems that each column of 
closed-shell atoms are slowly converging to the alkali earth case, 
migrating slowly from the  
low-energy edge of the exchange energy oscillations to the high-energy.
  
This last feature can be brought ought more plainly by
considering trends 
versus quantum number rather than $Z^{1/3}$.
We first note that closed-shell columns with the same value of the combination 
$n\HO+l\HO$, involving the principle quantum number and total angular momentum quantum 
number of the highest occupied atomic orbital, are located at a fixed distance from each 
other in the periodic table.  
Thus, the $n\HO$-s closed-shell atom is two columns from the closed-shell $(n\HO\!-\!1)$-p 
atom, which is six columns from $(n\HO\!-\!2)$d, etc.  
As $n\HO$ and $Z\to\infty$, the length of each row of the periodic table also 
diverges to infinity, and so these fixed differences become negigibly small fractions 
of an oscillation in Fig.~\ref{fig:bigZ_LDA_no_split}.
Thus $n\HO+l\HO$ yields a proxy for $Z^{1/3}$, one that is easier to construct 
a model for, and which allows a direct comparison between atoms in different 
colunns.

\begin{figure}[htb]
\includegraphics[trim=0cm 0cm 0cm 0cm,clip=true,width=1.0\linewidth]
{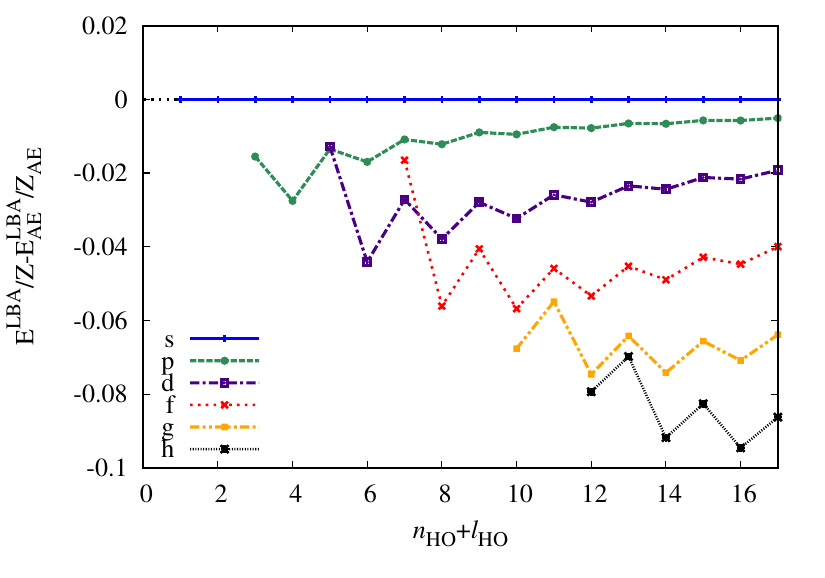}
\caption{
LDA exchange energy per electron for several closed-shell columns of the 
extended periodic table, measured 
relative to that of the nearest alkali earth atom.
Difference between LDA exchange energy per atom and that of 
He and the alkali earth
with the same $n\HO + l\HO$, where $n\HO$ is the principle quantum number and 
$l\HO$ is the angular momentum quantum number.  $p$ stands for closed $p$-shell
atoms (noble gases), etc.
}
\label{fig:bigZ_vs_n}
\end{figure}

As we can directly compare atoms across different columns of the periodic table which
have the same value of $n\HO + l\HO$, it is instructive to plot the exchange 
energy-per-electron for each closed-shell column
in our data set relative to that of the alkali earths, as shown in 
Fig.~\ref{fig:bigZ_vs_n}. 
This shows clear and reasonably smooth convergence of the
$p$ and $d$ series to the alkali earths, while the higher $l\HO$ columns seem to swerve
away for a few cycles before falling into line.
It seems likely that all
the columns will share the same asymptotic slope (i.e., leading order trend) in 
Fig.~\ref{fig:bigZ_LDA} as the alkali earths, but possibly have somewhat different 
asymptotes.  
The oscillatory
part of each column seems also to converge onto the alkali earth pattern, more slowly
for the higher $l\HO$ columns.

\ssec {Model for the large-$Z$ expansion of the alkali earth series}
The previous figures suggest a protocol for extracting the leading
order coefficient for the beyond-asymptotic exchange of the LDA.  As
$Z \to \infty$ or equivalently, $n\HO \to \infty$ for any closed-shell
column of the periodic table (fixed $l\HO$), we hypothesize the staircase
trend should gradually merge with that of the alkali earths and thus
the leading order for a given column should be given by that
of the alkali earths.  It is quite possible that lower-order terms
should still be $l$-dependent. 
From the visual analysis of the alkali earths, it seems apparent 
that the leading order term for beyond-TF LDA for this column should be of 
order $Z^{1/3}$ or equivalently, $n\HO + l\HO$ -- 
the straight-line behavior not seen in Fig.~3.  
In addition, as the oscillations between even and odd filled-shell atoms are 
exactly periodic vs principle quantum number $n\HO$, we can model then with
a simple oscillatory term.
As there is little evidence that the oscillation amplitude grows with $n\HO$
we posit the following extension of the large-$Z$ expansion, written
parametrically in terms of highest-occupied shell quantum numbers $n\HO(Z)$ and $l\HO(Z)$:
\ben
\label{eq:fitlda}
\begin{split}
E^{LBA}_{fit}/Z \to &\mathcal{A}\xi + \mathcal{B}\ln{(\xi)} + 
       \left [\mathcal{C} + \mathcal{\delta C}(-1)^\xi \right ]  \\
       & + \mathcal{D}/\xi + \mathcal{E}\ln{\xi}/\xi + \cdots
\end{split}
\een
where $
    \xi = n\HO + l\HO
$, so that $\xi=n\HO$ for Helium and the alkali earths.
Given the existence of a $\ln{Z}$ term for beyond-LDA exchange, 
we have added the possibility of a log term; otherwise the expression
is equivalent to an expansion in powers of $Z^{1/3}$.
We can also consider higher order terms of order $(n\HO + l\HO)^{-1}$, 
equivalent to $Z^{-1/3}$.

To ascertain the most likely form of the asymptotic expansion of
LDA exchange for the alkali
earth series,
we compare statistical fits to various models with terms down to
O(1) in $\xi$, an exercise similar to that carried out for beyond-local exchange 
in Ref.~\onlinecite{ARCB22}.
We first restrict our fits to values of $n\HO \geq 10$, restricting our focus to higher
$Z$ while keeping a reasonable number of data points.  
This should help maximize the footprint of the
leading terms in the LDA asymptotic series, and allow us to truncate higher-order terms.

\begin{figure}
	\includegraphics[trim=0cm 0cm 0cm 0cm,clip=true,width=1.0\linewidth]
	{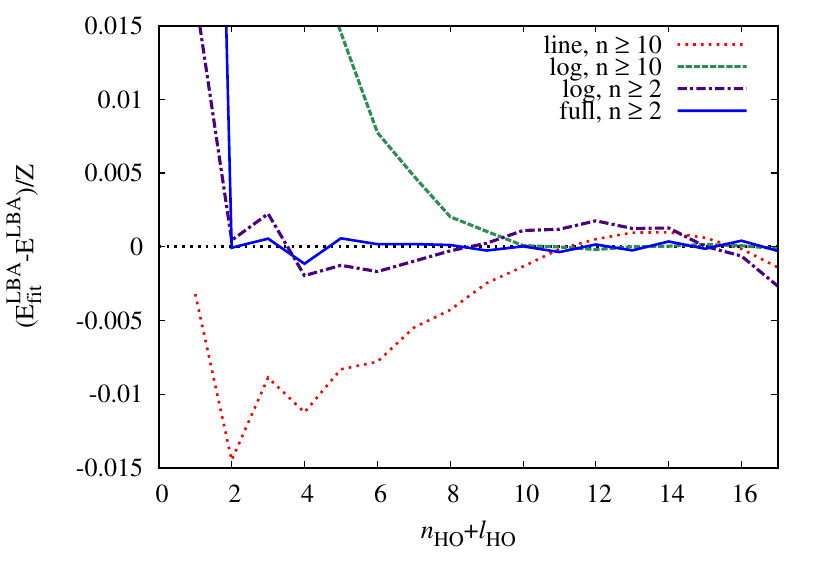}
	\caption{Difference of the LDA vs $n\HO$ for He and 
        alkali earth atoms and 
        various fits of the form Eq.~\ref{eq:fitlda}, as described in the text
        and in Table~\ref{table:fitLDAalt}.}
        
	\label{fig:LDAexpansion_diff}
\end{figure}

Since the trend in data for a given column is extremely smooth, 
the quality of a fit is not easily tested by usual statistical measures, so we 
judge by visual comparison, as shown in Fig.~\ref{fig:LDAexpansion_diff}. 
This shows the deviation of various least-square fits
of the LDA exchange energy per electron for the AE series.
These include a purely linear trend (labeled ``line"), with logarithmic 
coefficient $\mathcal{B}$ set
to zero in Eq.~(\ref{eq:fitlda}), and one with both linear and log trends (``log").  
Fit coefficients for these fits are 
Fits with $\mathcal{A}=0$ were clearly poorer and are not shown.
\READY{(More details on statistical fits and a complete list of models 
we considered are shown in supplemental information.}

Note that the error of the purely linear fit has noticeable curvature and thus 
the reported coefficient $\mathcal{A}=0.009$ or 9~mHA must be an underestimate 
of the slope. 
Adding the 
logarithmic term greatly improves the fit to the data, basically reproducing
the trend within numerical error for $n\HO \geq 10$, and suggesting a value for
the leading order coefficient of $\mathcal{A}=14$~mHa.
\begin{table*}[htb]
	\caption{Table of 
                fit parameters to the beyond-asymptotic contribution to the LDA
                exchange energy $E\x\LBA$ for closed-shell atoms segregated by 
                valence angular
		momentum $l\HO$.
		First column gives the angular momentum for highest occupied orbital
		Second column gives the minimum value of $\xi=n\HO+l\HO$ used in the 
		fit.  Missing coefficients indicate the coefficient has been fixed at zero.
                Fits are done by Levenberg-Marquardt method, with
		error in statistical fit given in parentheses.
}
	\begin{tabular}{ll|rrrrrr}
$l\HO$ & $n\HO$ &
$\mathcal{A}$ &
 $\mathcal{B}$ &
 $\mathcal{C}$ &
 $\mathcal{\delta C}$ &
 $\mathcal{D}$ & 
 $\mathcal{E}$ 
\\ \hline\hline
0 & $\geq$ 10 & 0.00925(18) &            & -0.0709(24)  & 0.0218(4)   & & \\ 
0 & $\geq$ 10 & 0.01433(31) & -0.068(4)  & 0.035(6)     & 0.02180(5)  & & \\ 
1 & $\geq$ 10 & 0.0134(5)    & -0.048(7)  & -0.010(11)   & 0.02147(9) & & \\ 
2 & $\geq$ 10 & 0.0124(15)  & -0.021(20)  & -0.082(32)  & 0.02049(27) & & \\ 
3 & $\geq$ 12 & 0.013(4)    & -0.016(6)    & -0.12(10)   & 0.0195(4) & & \\ \hline 
0 & $\geq$ 2 & 0.00944(32)  & -0.0087(24) & -0.0507(23) & 0.0221(4)   & & \\
0 & $\geq$ 2 & 0.0203(11)   & -0.282(30)  & 0.70(8)     & 0.02220(14) & -0.68(7) & -0.71(8)\\
\hline\hline
       \end{tabular}
\label{table:fitLDAalt}
\end{table*}

Table~\ref{table:fitLDAalt} shows coefficients of fits for
several columns of the periodic table in addition to the alkali earths -- corresponding
to filled $p$, $d$ and $f$ valence shells. 
While the coefficient $\mathcal{A}$ is somewhat sensitive to
the form of the asymptotic expansion taken, it is roughly independent of
column.  
Other coefficients are understandably less reliable, except for the amplitude 
of the even-row to odd-row oscillation.  This is quite stable for all columns 
of the table, all fit ranges and model forms.

Next, as a verification of the robustness of the fit parameters
we find, we repeat our calculation over a greater range of data, 
including all points with $n\HO \geq 2$.  
Fit 6 of Table~\ref{table:fitLDAalt} 
(``log, $n\geq 2$" in Fig.~\ref{fig:LDAexpansion_diff}) is
the same functional form as fit 2 in the table, but now it is seen 
to 
be unable to capture the observed trend at large $Z$.  To do so, we need
to add two additional sub-dominant terms in the series: 
$\mathcal{D}/\xi + \mathcal{E}\ln{\xi}/\xi$.  These results are shown as 
fit 7 in Table~\ref{table:fitLDAalt} and labeled as 
``full" in 
Fig.~\ref{fig:LDAexpansion_diff}, which shows that the resulting error is very
nearly reduced to small fluctuations for all rows except $n\HO=1$.
The penalty is that now, the leading order coefficient has changed yet again,
to $\mathcal{A} = 20$~mHa.

This result raises an essential problem limiting further progress -- the 
LDA exchange is inherently much more complicated than the beyond-LDA series.
In addition to the many terms needed, the presence of both logarithmic
and power law terms means that an accurate asymptotic trend will be hard to determine
even with the extended range of data we have for the LDA, and overfitting
is a clear danger.  Thus unfortunately, we do not expect our exercise to yield 
an unambiguous value for $\mathcal{A}$, and even less for $\mathcal{B}$ or other
subleading terms.

Nevertheless,
we should like our results for the alkali earths to be expressed
as an expansion in $n\HO$ converted into a conventional function of $Z$.
The $Z$ value for alkali earths for even rows of the periodic table (valence
shell configurations of 2s$^2$, 4s$^2$, 6s$^2$, etc.) is given exactly by
       \ben
       Z_{AE}(j)=\sum_{k=1}^j (2k)^2=\frac{2}{3}j(j+1)(2j+1)
        \label{eq:Zofj}
       \een
where the principle quantum number of the highest-occupied 
shell is $n\HO=2j$, and $j$ is an integer.  
\READY{
For any of the other series, take $j = (n\HO + l\HO)/2 (=\xi/2)$.  Then
\ben
Z_l(j) = Z_{AE}(j)-2 l\HO^2.
\een
}
\READY{We then consider $j$ as a continuous variable and invert $Z_{AE}(j)$ to
find, exactly for even-row alkali earths:}
	\ben
	\begin{split}
		&j(Z_{AE})+\frac{1}{2}=\\
		&+\bigg\{\left(\frac{3Z_{AE}}{8}\right)
		+\left[\left(\frac{3Z_{AE}}{8}\right)^2
		-\left( \frac{1}{12}\right)^3\right]^{1/2}\bigg\}^{1/3}\\
		&+\bigg\{\left(\frac{3Z_{AE}}{8}\right)
		-\left[\left(\frac{3Z_{AE}}{8}\right)^2
		-\left( \frac{1}{12}\right)^3\right]^{1/2}\bigg\}^{1/3}
	\end{split}
        \label{eq:jofZ}
	\een
\READY{
Now this expression yields an asymptotic expansion for $j$ in $Z$.  
In fact, Eq.~\ref{eq:Zofj} matches that for the summation of the 
total kinetic energy of a hydrogenic or Bohr atom}
(one with noninteracting particles feeling a central Coulomb potential.)
From this, one can derive~\cite{BCGP16} the following expansion:
       \ben
       \label{eq:Zexpans}
       j^{even}_{AE}(Z)=z -\half +\frac{1}{12z}+...
       \een
where $
        z=\left({3Z}/{4}\right)^{1/3}$.
For even-row atoms, $E_{fit}^{LBA}/Z$ then becomes
       \ben
       \begin{split}
       	\label{eq:nologexpansionZ}
              E_{fit}^{LBA}/Z\to 2\mathcal{A}z+\mathcal{B}\ln(2(z-1/2+(12z)^{-1}))\\
              (\mathcal{C}+\delta\mathcal{C}-\mathcal{A})+2\mathcal{A}(12z)^{-1}
       \end{split}
       \een
\READY{
Expanding the log term to order $z^{-1}$ and converting $z$ to $Z$ 
produces the expression}  
       \ben
              E_{fit}^{LBA}/Z\approx A\x\LDA Z^{1/3}+B\x\LDA\ln(Z)+C\x\LDA+D\x\LDA Z^{-1/3}
       \label{eq:fitLBA}
       \een
where $A\x\LDA = 6^{1/3}\mathcal{A}$.
For the two best fits to the alkali earths discussed in the text -- 
fit 2 and fit 7 of Table~\ref{table:fitLDAalt} 
we end up $A\x\LDA=25$ and $36$~mHa respectively.  
For odd-row alkali earths, the appropriate asymptotic expansion is
that with $\mathcal{\delta C}$ replaced by $-\mathcal{\delta C}$
   in Eq.~\ref{eq:nologexpansionZ}.  \READY{But in addition, one has to alter
the asymptotic map from $Z$ to $j$ to 
\ben
       j^{odd}_{AE}(Z)=z -\half -\frac{1}{6z}+...
\een
Inspection shows that if we use this expression to determine the 
odd rows, the leading order terms in Eq.~(\ref{eq:fitLBA}) are unchanged, but
terms starting with $D\x\LDA$ will change dramatically.
This is a small discrepancy but will
complicate the formulation of the column-dependent correction to our fit.
}

\begin{figure}[!htbp]\centering				    			  
	\includegraphics[width=1.0\linewidth,height=0.41\textheight,keepaspectratio]{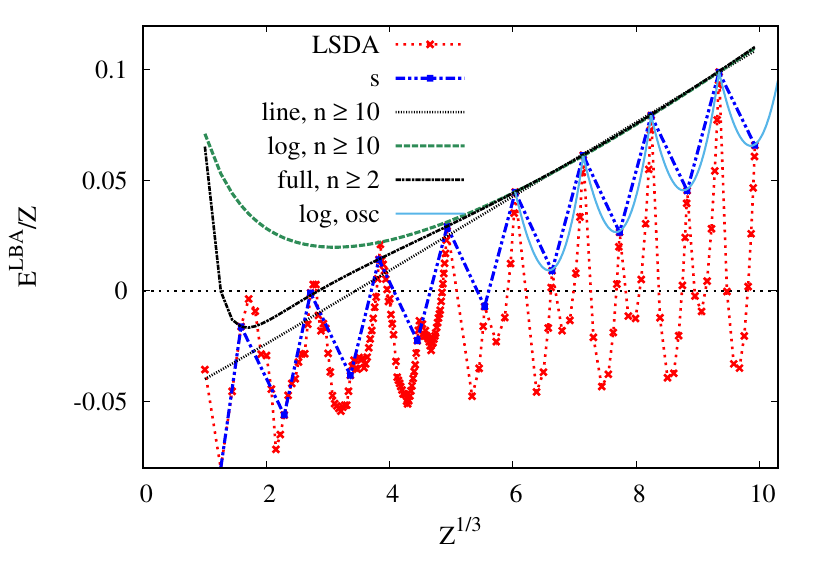}
	\caption{\label{fig:ExpansionZ}
		 The beyond asymptotic component of the LSDA ($E\LBA$) plotted vs $Z^{1/3}$ (red) with He and the alkali earths highlighted (blue); along with various fits parametrized in Table~\ref{table:fitLDAalt} discussed in the text.  The LSDA data for filled s-shell atoms is highlighted in blue. Fits shown are fit 1 of Table~\ref{table:fitLDAalt} (``line, $n\geq10$"), fit 2 (``log, $n\geq10$"), and fit 7 (``full, $n\geq2$").  Light blue shows the oscillatory 
extension of fit 2 discussed in Sec.~\ref{sec:oscillatory}.
}
\end{figure}

Figure~\ref{fig:ExpansionZ} is a reprise of Fig. \ref{fig:bigZ_LDA_no_split} showing various fits to the even $n\HO$ filled s-shell series of the form of Eq.~\ref{eq:nologexpansionZ}.
The basic linear fit, fit 1 of Table \ref{table:fitLDAalt}, already captures the rough trend of the data; the remaining error is reasonably small.
The second fit of table \ref{table:fitLDAalt} which includes a logarithmic term (non-zero $\mathcal{B}$), seems to capture the large Z asymptotics particularly well, given both figures~\ref{fig:LDAexpansion_diff} and ~\ref{fig:ExpansionZ}, but diverges for small Z.  
When including lower shell data in fits, complicated transient features become apparent, for example the region in between $2\geq Z^{1/3} \leq 6$.  These are fit well by fit 7 of Table~\ref{table:fitLDAalt}, 
(``full, $n\geq2$" in Fig.~\ref{fig:bigZ_LDA_no_split}), 
with six fitting parameters.  
These, however, may be the result of higher order oscillatory terms not captured well by the single oscillatory term $\delta\mathcal{C}$.
We also note that this fit is not nearly as good as fit 2 for large $Z$.  
This suggests that fit 2 gives the best predictions 
of $\mathcal{A}$ and $\mathcal{B}$ for the large Z asymptotic expansion.

\ssec{Oscillatory terms}
\label{sec:oscillatory}
\READY{
We finish with a short discussion of what issues are involved
with characterizing the full oscillatory pattern in the beyond-asymptotic
LDA.
From Fig.~\ref{fig:bigZ_LDA_no_split} and~\ref{fig:bigZ_LDA},
 we have a period that takes up two rows
of the periodic table, broken up into an even-row and odd-row sub-pattern.
To characterize this pattern, we construct a variable $\nu(Z)$ 
describing the partial occupation over a two-row period 
which varies uniformly between zero and one 
between
two even-row alkali earths, taking the value 1/2 at the odd-row alkali earth.
Similarly, one may need a variable $\lambda(Z)$ that does the same on a
per-row basis.
The function $j^{even}_{AE}(Z)$ that we used to convert our smooth fit in 
$n\HO$ to the alkali earths may be converted to a function of $Z$ usable
over the entire periodic table using an asymptotic expansion that depends
on both $Z$ and $\nu$:\cite{E88,BCGP16}
\ben
      j(Z,\nu) = z - \frac{1}{2} + \frac{1 - 12\nu(1-\nu)}{12z}
      \label{eq:monolith}
\een
With this formula, we can construct a single model for all the 
alkali earths, including oscillatory effects, as
\ben
      E\LBA/Z \to \mathcal{A}2j(Z,\nu) + \mathcal{B}\ln{[2j(Z,\nu)]}
           + \mathcal{C} + \delta\mathcal{C}[1 - 8\nu(1-\nu)].
\een
In the last term, we include an $\nu$-dependence similar to that of 
Eq.~(\ref{eq:monolith}).  
This is shown in Fig.~11 in light blue,
using fit 2 of Table~II, for rows with $n\HO$
greater than 10.  The agreement for all alkali earths is quite good.
}

\READY{
A full model of the entire periodic table will require a 
good model for the low-energy edge of the oscillatory pattern.
This lower edge has limited data 
for which our calculations prove to be less numerically reliable.
Preliminary work suggests that this could be fit by a 
logarithmic term shifted by a constant -- that is, with no $Z^{1/3}$ term.
}

\sec {Discussion}
\label{sec:outcomes}

This paper follows up on a report~\cite{ARCB22} and analysis of the conjectured
$Z\ln{Z}$ term in the large-$Z$ asymptotic expansion of the beyond-LDA 
contribution to exchange energies of neutral atoms.  A simple model for this
contribution on a per-electron basis is thus $\Delta E\x/Z \sim  \Delta B\x\ln{Z} + \Delta C\x$ 
where $\Delta B\x$ is taken to be 3/4 that derived from the Bohr atom and $\Delta C\x$ is the 
beyond-LDA exchange energy of the hydrogen atom.
This paper analyzes to what extent these beyond-LDA results are amenable to 
analysis by popular forms of density functional theory such as
GGAs and meta-GGAs and also  
describes the LDA exchange energies themselves, particularly 
determining a $Z^{4/3}$ behavior for the leading correction to the local limit.  These
studies demonstrate significant complications beyond what 
can be predicted by tools such as dimensional analysis and models such as the 
gradient expansion. 

Some perspective on these results may be obtained by reconsidering
the Lieb-Simon scaling that we began our story with. This process is a 
scaling of the potential, but there also exists dual procedure of density scaling:
\ben
\n_\zeta(\br) = \zeta^2\, \n(\zeta^{1/3} \br).
\een
which differs from the usual coordinate scaling in DFT, as the number of particles changes.
Within TF theory, the scaled density is the ground-state of the scaled potential, but
this is not true more generally.

Early work analyzing the asymptotic form of exchange\cite{PCSB06,EB09}
took the approach of density scaling, and the exact density weakly
approaches the TF density in the limit of large $Z$.   (Weakly here means
sufficiently smooth integrals over the exact density approach zero relative
error when evaluated with the TF density.)  Moreover, the relative error in the TF
total energy also vanishes.  Assuming (incorrectly) that 
differences between the exact and approximate densities would
contribute mainly to higher orders and to 
oscillations in the exchange energy characteristic of the LDA, 
suggests
\ben
    E\x \to  -d_0 Z^{5/3} - \Delta C\x Z + \cdots
\een
for exchange.  In fact, the 
\READY{beyond-LDA} data was tested to include a $Z^{4/3}$ term, 
but its coefficient
was found to be numerically indistinguishable from zero\cite{EB09}. 

\READY
{The LS theorem guarantees weak convergence to the TF limit for the total
energy and the results of Schwinger\cite{S81} and Conlon\cite{C83} imply
a similar principle for the leading contribution to the exchange energy.
However, beyond such dominant terms, whether or not the gradient expansion
can yield the next subdominant correction is a delicate question.  The
gradient expansion for the (non-interacting) kinetic energy {\em does}
yield the exact contribution to the $Z^{5/3}$ in the total energy, without
modification.}  
\READY{Table~I shows that the true
gradient expansion limit, implemented by PBEsol gets $\Delta C\x$ closer 
than any other density functional, 
(off by 20\% at most).  Our estimate for the fourth-order gradient 
expansion contribution to this quantity, about 8~mHa, almost exactly makes up 
this deficit.  
What the gradient expansion does not predict properly, either 
for kinetic energy or for exchange is an 
additional anomalous term not predictable from density scaling -- 
the Scott term of order $Z^2$ for the former and the $Z\ln{Z}$ term
for exchange.}

A contrast could be made to the situation with quantum dots\cite{KR10},
whose asymptotic series for the total energy and for exchange each have the 
same form as that generated by density scaling.
It remains an open question what should be the necessary
conditions in a scaled potential for anomalous powers to appear in the 
associated asymptotic series.

       \ssec{Semilocal approximations}
The analysis of semilocal density functionals
adds nuance to our prior conclusions.  
First, 
the logarithmic term for the exchange energy-per-particle in such a model is 
uniquely determined by its second-order gradient expansion in the limit 
of a slowly varying density, and the coefficient, $\Delta B\x$,
by the coefficient $\mu$ of this correction.  
But also the second coefficient $\Delta C\x$ is roughly linear in $\mu$ as well.  
A GGA thus has the task of attempting to fit two asymptotic coefficients with 
one adjustable parameter.
This task is made more difficult in that the value of $\mu$, $1/3$,  needed to obtain the 
$B$ coefficient of our model, is significantly larger than any encountered in a successful 
GGA.
What a successful GGA like PBE or BLYP does is find a useful
compromise between obtaining an accurate value for $\Delta B\x$ and one for $\Delta C\x$
by choosing a $\mu$ somewhere between the correct value and that needed to match 
asympototics of the large-$Z$ atom.  \READY{In contrast, as discussed above,
the implementation of the exact gradient expansion limit matches $\Delta C\x$
reasonably while introducing large errors in $\Delta B\x$.}

In contrast, metaGGAs have the flexibility to match both $\Delta B\x$ and $\Delta C\x$ 
because they incorporate explicit control over the second-order and
fourth-order gradient expansion corrections to the LDA.  But doing so likely produces
coefficients quite far from those optimal for a 
slowly-varying periodic system.  It is uncertain whether this would be a profitable
thing to do.  

But all these conclusions are limited by the limited range of our data and 
capacity of our theoretical methods to determine the exact asymptotic behavior of
atoms.  We have seen that one could generate an alternate model supposing B88
exchange to be the correct picture, with behavior for $Z<100$ determined by a large
number of subdominant terms in an asymptotic expansion while the leading terms 
only becoming dominant for $Z \gg 100$.  The need for such subdominant terms to
describe B88's asymptotic behavior is likely due to its complicated form as a function
of $s^2$.  We cannot prove that they are not there for exact exchange.
\READY{Rather, our conclusions stand on the surprising accuracy of our 
simple two parameter fit, consistent with the expectation
that asymptotic expansions are efficient.}

       \ssec{ Outcomes for LDA }
A full picture of the asymptotic expansion of exchange involves the description of
LDA exchange beyond the TF approximation.
The result is a rich and complex oscillatory pattern showing the full complexity
of the periodic table and which we only start to unravel.

Because of the need to consider logarithmic terms in our modeling, 
we are unable to get a definitive fit, \READY{even for a smooth average trend,}
in the fashion of the beyond-LDA data.
This awaits accurate accurate exchange data at much larger values of $Z$.
But we are able to make a number of conclusions -- primarily that the leading beyond-TF
term in exchange of neutral atoms  is of order $Z^{4/3}$.  The value of 
the coefficient depends on the fit strategy used but is likely between 10 to 20~mHa.
This estimate is made by fitting the data for the alkali earth column and 
noting that the 
energy series of any other closed-shell column approaches that of the 
alkali earths at least to this order.

The overall trend of oscillations is quite complex.  The pattern
(involving closed-shell atoms only) depends on the fraction of 
filling of each row of the periodic
table superimposed on a trend that takes two rows to repeat.
At least for closed-shell systems, the local maxima of this two-row
pattern seem to be the even-row and odd-row alkali earths.  
For this and likely any column of the periodic table there is a 
distinct oscillation 
between even and odd rows of the periodic table with an asymptotic form of 
$\pm\mathcal{\delta C}Z$, for $\mathcal{\delta C} = 22$mHa.  
The lower boundary is formed by a series of atoms with near minimum 
$n\HO-l\HO$ --
those that nearly maximize the angular momentum of the
valence shell for a given value of $n\HO + l\HO$.
The amplitude of the complete oscillatory pattern 
apparently grows roughly as $Z^{4/3}$, similar to
that seen by Englert for the total energy.

The cause of these effects is the 
appearance of shell structure changing the charge density from
the scale-invariant Thomas-Fermi density.  The very strong dependence of exchange 
on the angular momentum quantum number of the frontier shell, as well as common
sense consideration of energetics,
indicates that these changes should primarily involve the outer shells of 
the atom. 
The most notable density change is at large radial distqnces $r$ from the nucleus:
from the asymptotic of $1/r^6$ for TF to an exponential decay at finite radius.  
However, there can be a 
non-negligible contribution from a large number of outer shells of the atom, 
as has been seen in a somewhat different context in Ref.~\cite{RC21}.  

One argument why alkali earths should present the ``roof" for exchange energy comes from 
considering the exchange contribution of the outermost shell.  
The exchange energy can be expressed as the expectation of the interaction of an electron 
with its exchange hole -- the change in overall electron density elsewhere given the
particle is observed at some particular point in space.
If this electron is in a valence $s$-shell, its exchange hole will
be diffused over the entire valence region, leading to a relatively
weak exchange energy.  This could very well be positive relative to 
that predicted by Thomas-Fermi-Dirac theory.  For a frontier shell 
with a large degree of degeneracy, greater localization of the exchange
hole is possible and thus a lower exchange energy.

Although they may not impact development of 
KS DFT, these issues in the LDA and beyond-LDA energy for exchange 
illustrate the 
difficulty of achieving accurate orbital-free 
models of the energy and its XC component, and point to the 
large amount of work that still remains to understand the large-$Z$ limit 
of atoms.

\section*{Supplementary Material}
See supplementary material for tables of LSDA and GGA exchange energies, numerical stress tests, additional fit data and additional plots of exchange energies. 
\READY{
For OEP data, see the supplementary material of Ref.~\onlinecite{ARCB22}.}
Additional data is available on request.

\begin{acknowledgments}
A.C. would like to thank Jian-Wei Sun for useful conversation, 
and Eberhard Engel for use of his atomic DFT code, OPMKS.
KB funded by NSF grant CHE-2154371.
\end{acknowledgments}

\appendix
\section{Large Z stress test}
	As a check on the validity of large $Z$ atomic structure
	calculations, we have compared the energetics from two independent
	atomic density functional codes, FHI98PP~\cite{FHI98PP} and 
	OPMKS~\cite{ED99}.  
	The former is done using the default integration grid with geometric growth
	factor $\gamma=0.0247$, and 
	and a finer grid $\gamma=0.123$ as described in the text. 
	Table~\ref{table:HOAOfine} shows calculations for the energy eigenvalue of the highest-occupied atomic orbital (HO) of helium and noble gas atoms up to $Z=976$. 
	These prove to be an energy measure sensitive to numerical issues such as failure to converge to a solution in FHI98PP.  PW91 exchange and correlation was used for these calculations.
	\begin{table}[htb]
		\caption{\label{table:HOAOfine} Comparison of the
			HO eigenvalue calculated on a coarse and fine grid using FHI98PP and using OPMKS}
		\begin{ruledtabular}
			\begin{tabular}{lccc}
				Z & HO fine grid & HO coarse & OPMKS \\ \hline
				2 & -0.57025586 & -0.57025588 & -0.57025583 \\ \hline
				10 & -0.49784696 & -0.49784690 & -0.49784691 \\ \hline
				18 & -0.38222044 & -0.38222035 & -0.38222041 \\ \hline
				36 & -0.34625570 & -0.34625554 & -0.34625569 \\ \hline
				54 & -0.30977911 & -0.30977894 & -0.30977910 \\ \hline
				86 & -0.29313618 & -0.29313608 & -0.29313616 \\ \hline
				118 & -0.27392048 & -0.27392058 & -0.27392045 \\ \hline
				168 & -0.26363740 & -0.26363782 & -0.26363735 \\ \hline
				218 & -0.25147948 & -0.25148038 & -0.25147939 \\ \hline
				290 & -0.24426230 & -0.24426386 & -0.24426218 \\ \hline
				362 & -0.23575844 & -0.23576087 & -0.23575826 \\ \hline
				460 & -0.23032261 & -0.23032611 & 			  \\ \hline
				558 & -0.22397950 & -0.22398436 & -0.22397916 \\ \hline
				686 & -0.21969293 & -0.21969939 & 			  \\ \hline
				814 & -0.21474382 & -0.21475225 & 		     \\ \hline
				976 & -0.21125111 & -0.21126178 & 			   \\ 
			\end{tabular}
		\end{ruledtabular}
	\end{table}
\section{Energy Data}
	Tables ~\ref{table:Ex},\ref{table:Ex1},\ref{table:Ex2},\ref{table:Ex3} give total exchange energies for LSDA exchange calculated by OPMKS for atoms $Z\leq120$ as well as the LDA for closed shell atoms $Z>120$ (same as LSDA) for and various GGAs calculated by FHI98PP.  GGAs are not spin polarized calculations.  An entry of "nan" indicates that the calculation did not converge.
		\begin{table}[htb]
			\caption{\label{table:Ex} Exchange energies as calculated by FHI98PP in non-relativistic mode using the exchange functional stated, with the exception of the LSDA which was calculated by OPMKS for atoms up to 120, and FHI98PP for higher Z atoms.} 
			
			\begin{tabular}{l|rrrr}
				Z & PBEsol & PBE & BLYP & LSDA \\ \hline
				1 & -0.285103 & -0.301759 & -0.305921 & -0.256426 \\ \hline 
				2 & -0.952651 & -1.005099 & -1.018337 & -0.861740 \\ \hline 
				3 & -1.665161 & -1.751734 & -1.771358 & -1.514295 \\ \hline 
				4 & -2.507204 & -2.633584 & -2.657785 & -2.290333 \\ \hline 
				5 & -3.527610 & -3.695906 & -3.726510 & -3.246785 \\ \hline 
				6 & -4.776838 & -4.988579 & -5.027871 & -4.430191 \\ \hline 
				7 & -6.272321 & -6.529170 & -6.578122 & -5.857031 \\ \hline 
				8 & -7.790568 & -8.100493 & -8.154102 & -7.300277 \\ \hline 
				9 & -9.564857 & -9.928070 & -9.989481 & -8.999038 \\ \hline 
				10 & -11.610176 & -12.027537 & -12.099347 & -10.966746 \\ \hline 
				11 & -13.451905 & -13.924267 & -14.006053 & -12.729476 \\ \hline 
				12 & -15.366169 & -15.896179 & -15.986368 & -14.563428 \\ \hline 
				13 & -17.368448 & -17.953607 & -18.053264 & -16.485917 \\ \hline 
				14 & -19.507646 & -20.148558 & -20.259722 & -18.544496 \\ \hline 
				15 & -21.788564 & -22.486058 & -22.609315 & -20.743147 \\ \hline 
				16 & -24.076698 & -24.832382 & -24.966611 & -22.949589 \\ \hline 
				17 & -26.515780 & -27.330118 & -27.475697 & -25.305450 \\ \hline 
				18 & -29.107513 & -29.981392 & -30.139415 & -27.812161 \\ \hline 
				19 & -31.540945 & -32.475933 & -32.646217 & -30.162201 \\ \hline 
				20 & -34.021221 & -35.018503 & -35.200554 & -32.559071 \\ \hline 
				21 & -36.780064 & -37.837632 & -38.033415 & -35.236844 \\ \hline 
				22 & -39.706642 & -40.824452 & -41.034752 & -38.080882 \\ \hline 
				23 &    nan     &    nan     & -44.438807 & -41.332351 \\ \hline 
				24 & -46.317208 & -47.549392 & -47.796449 & -44.530068 \\ \hline 
				25 & -49.449931 & -50.749570 & -51.005773 & -47.571047 \\ \hline 
				26 & -52.794474 & -54.158130 & -54.429019 & -50.832824 \\ \hline 
				27 & -56.468921 & -57.893017 & -58.177009 & -54.423184 \\ \hline 
				28 & -60.172474 & -61.659715 & -61.959412 & -58.040636 \\ \hline 
				29 & -64.040916 & -65.591487 & -65.907619 & -61.822122 \\ \hline 
				30 & -67.795918 & -69.413739 & -69.750274 & -65.492277 \\ \hline 
				31 & -71.561204 & -73.242822 & -73.596713 & -69.168712 \\ \hline 
				32 & -75.407020 & -77.151814 & -77.524221 & -72.925515 \\ \hline 
				33 & -79.340876 & -81.149203 & -81.539986 & -76.769192 \\ \hline 
				34 & -83.245939 & -85.119973 & -85.529139 & -80.587205 \\ \hline 
				35 & -87.256845 & -89.195559 & -89.622997 & -84.509595 \\ \hline 
				36 & -91.373178 & -93.376894 & -93.822814 & -88.535671 \\ \hline 
				37 & -95.330802 & -97.402192 & -97.866120 & -92.403904 \\ \hline 
				38 & -99.315762 & -101.455560 &	 -101.937255 & -96.299795  \\ \hline 
				39 & -103.476697 & -105.682558 & -106.182519 & -100.371730  \\ \hline 
				40 & -107.763775 & -110.036018 & -110.554242 & -104.567434  \\ \hline 
				41 & -112.378509 & -114.712770 & -115.251191 & -109.089494  \\ \hline 
				42 & -116.931623 &    nan      & -119.890641 & -113.548978  \\ \hline 
				43 & -121.317441 & -123.792217 & -124.365738 & -117.839362  \\ \hline 
				44 & -126.040253 & -128.580983 & -129.170095 & -122.470059  \\ \hline 
				45 & -130.769236 & -133.378702 & -133.985919 & -127.105429  \\ \hline 
				46 & -135.773758 & -138.450985 & -139.073656 & -132.012189  \\ \hline 
				47 & -140.574957 & -143.322635 & -143.967816 & -136.721135  \\ \hline 
				48 & -145.380020 & -148.199147 & -148.867528 & -141.436736  \\ \hline 
				49 & -150.190853 & -153.078451 & -153.768105 & -146.155034  \\ \hline 
				50 & -155.069722 & -158.024503 & -158.736018 & -150.941087  \\ \hline 

			\end{tabular}
		\end{table}
	\begin{table}[htb]
		\caption{\label{table:Ex1} Continuation of table~\ref{table:Ex}} 
		
		\begin{tabular}{l|rrrr}
Z & PBEsol & PBE & BLYP & LSDA \\ \hline				
51 & -160.018061 & -163.040081 & -163.772972 & -155.795528  \\ \hline 
52 & -164.934009 & -168.025544 & -168.779985 & -160.621632  \\ \hline 
53 & -169.931908 & -173.091264 & -173.867222 & -165.528358  \\ \hline 
54 & -175.009592 & -178.236785 & -179.034156 & -170.513218  \\ \hline 
55 & -179.942326 & -183.240451 & -184.058395 & -175.353399  \\ \hline 
56 & -184.894972 & -188.264447 & -189.103066 & -180.213928  \\ \hline 
57 & -190.004274 & -193.443095 & -194.302285 & -185.230322  \\ \hline 
58 & -195.966874 & -199.472708 & -200.354797 & -191.103880  \\ \hline 
59 & -201.704369 & -205.278758 & -206.182989 & -196.749306  \\ \hline 
60 & -207.567828 & -211.210992 & -212.137415 & -202.520151  \\ \hline 
61 & -213.555438 & -217.267608 & -218.216238 & -208.414629  \\ \hline 
62 & -219.666426 & -223.447845 & -224.418680 & -214.431982  \\ \hline 
63 & -225.900563 & -229.751481 & -230.744510 & -220.571984  \\ \hline 
64 & -231.775153 & -235.696001 & -236.712778 & -226.355247  \\ \hline 
65 & -238.252463 & -242.241569 & -243.281721 & -232.742716  \\ \hline 
66 & -244.609510 & -248.667943 & -249.731623 & -239.008105  \\ \hline 
67 & -251.088082 & -255.216058 & -256.303258 & -245.394357  \\ \hline 
68 & nan         &	 nan   & nan         & nan          \\ \hline 
69 & -264.412667 & -268.680483 & -269.814727 & -258.532392  \\ \hline 
70 & -271.260166 & -275.598322 & -276.756100 & -265.285690  \\ \hline 
71 & -277.771794 & -282.181033 & -283.364169 & -271.704359  \\ \hline 
72 & -284.348360 & -288.828294 & -290.035553 & -278.185500  \\ \hline 
73 & -290.992645 & -295.543525 & -296.774193 & -284.732719  \\ \hline 
74 & -297.708352 & -302.330639 & -303.584266 & -291.350064  \\ \hline 
75 & -304.498753 & -309.192979 & -310.469278 & -298.040989  \\ \hline 
76 & -311.211928 & -315.978919 & -317.279032 & -304.661333  \\ \hline 
77 & -318.009788 & -322.848525 & -324.172348 & -311.364319  \\ \hline 
78 & -325.080531 & -329.988547 & -331.331978 & -318.334025  \\ \hline 
79 & -332.076152 & -337.056848 & -338.423389 & -325.230195  \\ \hline 
80 & -338.913998 & -343.969612 & -345.363313 & -331.974258  \\ \hline 
81 & -345.739757 & -350.867933 & -352.286930 & -338.703239  \\ \hline 
82 & -352.617522 & -357.816687 & -359.261580 & -345.483911  \\ \hline 
83 & -359.547982 & -364.818090 & -366.288284 & -352.316191  \\ \hline 
84 & -366.436336 & -371.779877 & -373.275519 & -359.110651  \\ \hline 
85 & -373.389574 & -378.804549 & -380.325715 & -365.968805  \\ \hline 
86 & -380.405332 & -385.891604 & -387.438058 & -372.887890  \\ \hline 
87 & -387.276708 & -392.837681 & -394.408340 & -379.662436  \\ \hline 
88 & -394.159238 & -399.795193 & -401.390311 & -386.448625  \\ \hline 
89 & -401.172455 & -406.881313 & -408.500740 & -393.364763  \\ \hline 
90 & -408.257414 & -414.039285 & -415.682685 & -400.350505  \\ \hline 
91 & -415.962771 & -421.816489 & -423.485682 & -407.960738  \\ \hline 
92 & -423.511974 & -429.438950 & -431.132767 & -415.411065  \\ \hline 
93 & -431.160424 & -437.161234 & -438.879571 & -422.959664  \\ \hline 
94 & -439.263323 & -445.339853 & -447.080632 & -430.955907  \\ \hline 
95 & -447.131841 & -453.283617 & -455.048726 & -438.722078  \\ \hline 
96 & -454.691592 & -460.917180 & -462.708893 & -446.186310  \\ \hline 
97 & -462.776870 & -469.075474 & -470.892098 & -454.172626  \\ \hline 
98 & -470.744149 & -477.116833 & -478.958906 & -462.040969  \\ \hline 
99 & -478.807734 & -485.254994 & -487.122420 & -470.004641  \\ \hline 
100 & -486.967635 & -493.489973 & -495.382716 & -478.063747 \\ \hline 
\end{tabular}
\end{table}
\begin{table}[htb]
	\caption{\label{table:Ex2}Continuation of table~\ref{table:Ex}} 
	
	\begin{tabular}{l|rrrr}
Z & PBEsol & PBE & BLYP & LSDA \\ \hline
101 & -495.223960 & -501.821863 & -503.739936 & -486.218473 \\ \hline 
102 & -503.576887 & -510.250829 & -512.194273 & -494.469059 \\ \hline 
103 & -511.446178 & -518.196149 & -520.164373 & -502.235361 \\ \hline 
104 & -519.635752 & -526.458925 & -528.456391 & -510.331367 \\ \hline 
105 & -527.758822 & -534.656210 & -536.679833 & -518.353806 \\ \hline 
106 & -535.943574 & -542.915471 & -544.964778 & -526.436787 \\ \hline 
107 & -544.190875 & -551.237690 & -553.312318 & -534.581323 \\ \hline 
108 & -552.363052 & -559.485586 & -561.586814 & -542.657967 \\ \hline 
109 & -560.605775 & -567.802772 & -569.930538 & -550.803254 \\ \hline 
110 & -568.917955 & -576.189296 & -578.343265 & -559.016307 \\ \hline 
111 & -577.299204 & -584.645196 & -586.825075 & -567.296985 \\ \hline 
112 & -585.749514 & -593.170631 & -595.376208 & -575.645463 \\ \hline 
113 & -593.980585 & -601.476876 & -603.710328 & -583.777322 \\ \hline 
114 & -602.257346 & -609.827158 & -612.088912 & -591.954413 \\ \hline 
115 & -610.579172 & -618.222415 & -620.511799 & -600.175466 \\ \hline 
116 & -618.858735 & -626.577875 & -628.895045 & -608.358631 \\ \hline 
117 & -627.193577 & -634.986538 & -637.331588 & -616.595915 \\ \hline 
118 & -635.580919 & -643.447490 & -645.820102 & -624.884164 \\ \hline 
119 & -643.831869 & -651.775518 & -654.174451 & -633.035711 \\ \hline 
120 & -652.090902 & -660.111784 & -662.537449 & -641.195821 \\ \hline 
152 &      nan    &    nan      &      nan    & -963.24472  \\ \hline
162 & -1083.90933 & -1095.11339 & -1098.72977 & -1068.78028 \\ \hline 
168 & -1147.97249 & -1159.63834 & -1163.43702 & -1132.23151 \\ \hline 
170 & -1169.18675 & -1181.01236 & -1184.86899 & -1153.24075 \\ \hline 
188 & -1384.59708 & -1397.84818 & -1402.22911 & -1366.74978 \\ \hline 
202 & -1557.45197 & -1571.80514 & -1576.61687 & -1538.13741 \\ \hline 
212 & -1682.21957 & -1697.36592 & -1702.48740 & -1661.85723 \\ \hline 
218 & -1757.04245 & -1772.66219 & -1777.97727 & -1736.05551 \\ \hline 
220 & -1781.83402 & -1797.61724 & -1802.99384 & -1760.63772 \\ \hline 
260 & -2376.19435 & -2395.16288 & -2401.80899 & -2350.77803 \\ \hline 
274 & -2589.51881 & -2609.61454 & -2616.71921 & -2562.60478 \\ \hline 
284 & -2742.42429 & -2763.33352 & -2770.76685 & -2714.43811 \\ \hline 
290 & -2833.92454 & -2855.32006 & -2862.95786 & -2805.29803 \\ \hline 
292 & -2864.25370 & -2885.81704 & -2893.51987 & -2835.41239 \\ \hline 
314 & -3240.26344 & -3263.65782 & -3272.07848 & -3208.99340 \\ \hline 
332 & -3552.85191 & -3577.73436 & -3586.75994 & -3519.60902 \\ \hline 
346 & -3797.70737 & -3823.74208 & -3833.25001 & -3762.93835 \\ \hline 
356 & -3972.78324 & -3999.64774 & -4009.50028 & -3936.92316 \\ \hline 
362 & -4077.50032 & -4104.86074 & -4114.92699 & -4040.98896 \\ \hline 
364 & -4112.23126 & -4139.76261 & -4149.89687 & -4075.50139 \\ \hline 
390 &      nan    &    nan      &      nan    & -4605.39831 \\ \hline 
412 & -5097.33491 & -5128.86128 & -5140.65967 & -5055.32866 \\ \hline 
430 & -5469.85244 & -5502.89115 & -5515.32025 & -5425.84307 \\ \hline 
444 & -5760.25369 & -5794.46673 & -5807.39807 & -5714.69083 \\ \hline 
454 & -5967.44325 & -6002.50313 & -6015.79310 & -5920.76822 \\ \hline 
460 & -6091.30751 & -6126.87403 & -6140.38596 & -6043.96787 \\ \hline 
462 & -6132.40804 & -6168.14908 & -6181.73168 & -6084.84529 \\ \hline 
488 & -6735.70871 & -6773.70159 & -6788.20765 & -6685.16919 \\ \hline 
510 & -7248.40529 & -7288.29035 & -7303.59062 & -7195.36575 \\ \hline 
528 & -7669.19259 & -7710.61874 & -7726.57631 & -7614.11588 \\ \hline 
\end{tabular}
\end{table}
\begin{table}[htb]
\caption{\label{table:Ex3} Continuation of table~\ref{table:Ex}} 

\begin{tabular}{l|rrrr}
Z & PBEsol & PBE & BLYP & LSDA \\ \hline
542 & -7996.59712 & -8039.21793 & -8055.69704 & -7939.94341 \\ \hline 
552 & -8230.01525 & -8273.49669 & -8290.34771 & -8172.23292 \\ \hline 
558 & -8369.56749 & -8413.56379 & -8430.64452 & -8311.11092 \\ \hline 
560 & -8415.89591 & -8460.06941 & -8477.22332 & -8357.21298 \\ \hline 
590 & -9234.38168 & -9281.12695 & -9299.39917 & -9172.31421 \\ \hline 
616 & -9937.37621 & -9986.38409 & -10005.61560 & -9872.32126    \\ \hline 
638 & -10532.65533 & -10583.57795 & -10603.62717 & -10465.07172 \\ \hline 
656 & -11019.81029 & -11072.29698 & -11093.02306 & -10950.16042 \\ \hline 
670 & -11398.21910 & -11451.92007 & -11473.18306 & -11326.96761 \\ \hline 
680 & -11667.81044 & -11722.38685 & -11744.03273 & -11595.41165 \\ \hline 
686 & -11828.98898 & -11884.08946 & -11905.97160 & -11755.90438 \\ \hline 
688 & -11882.51809 & -11937.79884 & -11959.75631 & -11809.20317 \\ \hline 
718 & -12791.86384 & -12849.83042 & -12872.92584 & -12715.00661 \\ \hline 
744 & -13576.69738 & -13636.97556 & -13661.06586 & -13496.79268 \\ \hline 
766 & -14240.29550 & -14302.52054 & -14327.45705 & -14157.82352 \\ \hline 
784 & -14782.68823 & -14846.50057 & -14872.13589 & -14698.12228 \\ \hline 
798 & -15203.72967 & -15268.77376 & -15294.96212 & -15117.54153 \\ \hline 
808 & -15503.64709 & -15569.57865 & -15596.16114 & -15416.29696 \\ \hline 
814 & -15682.99103 & -15749.45385 & -15776.27923 & -15594.94651 \\ \hline 
816 & -15742.57501 & -15809.22045 & -15836.12330 & -15654.29729 \\ \hline 
850 & -16935.40359 & -17005.05550 & -17033.29559 & -16843.17349 \\ \hline 
880 & -17970.11068 & -18042.45548 & -18071.86437 & -17874.33172 \\ \hline 
906 & -18863.19591 & -18937.87275 & -18968.29738 & -18764.34310 \\ \hline 
928 & -19617.28734 & -19693.93415 & -19725.22376 & -19515.83775 \\ \hline 
946 & -20232.98309 & -20311.23839 & -20343.24266 & -20129.41253 \\ \hline 
960 & -20710.64290 & -20790.14767 & -20822.71743 & -20605.42779 \\ \hline 
970 & -21050.83029 & -21131.23561 & -21164.20853 & -20944.43646 \\ \hline 
976 & -21254.28367 & -21335.22837 & -21368.44955 & -21147.18516 \\ \hline  
978 & -21321.89884 & -21403.02885 & -21436.32926 & -21214.56365 \\ \hline 
\end{tabular}
\end{table}

\section{Additional plots}
\begin{figure}
	\includegraphics[trim=0cm 0cm 0cm 0cm,clip=true,width=1.0\linewidth]
	{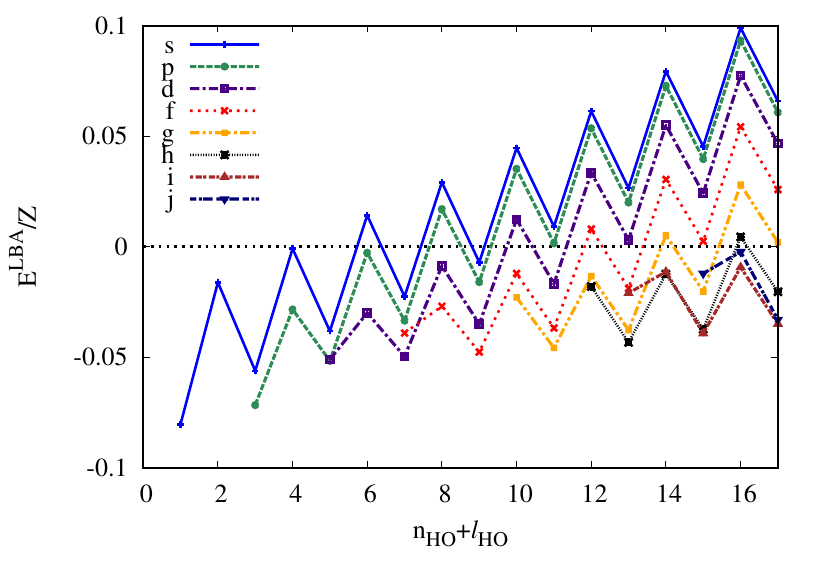}
	\caption{Plots the difference between KS LSDA exchange energies and TF LDA exchange vs principle quantum number plus the angular momentum quantum number of the last filled shell.}
	\label{fig:LDAvsN}
\end{figure}

Fig.~\ref{fig:LDAvsN} plots exchange-energy per charge split into columns of the periodic table, plotted vs the principle quantum number of the HO ($n_{HO}$)plus the angular momentum of the HO ($l_{HO}$).
\section{Statistical fits of beyond-asymptotic LDA data}
In Table~\ref{table:fit10} and~\ref{table:fit2} we present statistical fits
of a number of models for the beyond-asymptotic component of LDA exchange,
defined by Eq.~6 in the main text. The fits are to an expansion in $n_{HO}$, 
the principal quantum number of the highest-energy occupied orbital, given
by Eq.~33 of the main text, reprised, .  Shown in the SI are a number of fits 
using Levenberg-Marquardt nonlinear regression.  If a coefficient is missing,
then it has been constrained to be zero.  Standard errors are shown in parentheses, along with the $\chi^2_{red}$ measure, which is the $\chi^2$ measure of the fit divided by the number of independent data points in the fit.  A $\chi^2_{red}$ less than $1e-7$ was needed to reduce the error to numerical error rather than systematic error from a wrong fitting function.
\begin{table*}[htb]
\begin{tabular}{|rrrrrrl|}
\hline\hline
 $\mathcal{A}$ & $\mathcal{B}$ & $\mathcal{C}$ & $\mathcal{\delta C}$ & $\mathcal{D}$ & $\mathcal{E}$ & $\chi^2_{red}\times 10^{-6}$ \\
\hline
 0.00925(18) &         & -0.0710(24) & 0.02183(4) & & & 1.2\\
             & 0.122(7) & -0.261(17) & 0.0218(11) & & & 9.9\\
 0.01433(31) & -0.067(4) & 0.035(6)  & 0.02180(5) & & & 0.023\\
 0.01176(21) &           & -0.138(6) & 0.02179(7) & 0.434(36) & & 0.042\\
             & 0.309(12) & -0.93(4) & 0.02172(16) & 2.42(16) & & 0.21\\
 0.0209(14) & -0.242(36) & 0.48(9) & 0.021844(23) & -1.13(23) & & 0.003\\
 0.050(12) & -1.8(6) & 6.0(2.2) & 0.021842(14) & -0.53(28) & -10.(4) & 0.001\\
\hline\hline
\end{tabular}
\caption{Parameters for nonlinear fit beyond-TF LDA exchange data for alkali
earths to Eq.~{eq:beyondTF}.  All rows $n\geq 10$ are included in fitting set.  
Blank spaces indicate where a parameter is held at zero.  $\chi^2_{red}$ is
reduced $\chi^2$ measure, defined as the $\chi^2$ error of the fit divided
by the number of independent degrees of freedom in the fit set.}
\label{table:fit10}
\end{table*}

\begin{table*}[htb]
\begin{tabular}{|rrrrrrl|}
\hline\hline
 $\mathcal{A}$ & $\mathcal{B}$ & $\mathcal{C}$ & $\mathcal{\delta C}$ & $\mathcal{D}$ & $\mathcal{E}$ & $\chi^2_{red}\times 10^{-6}$ \\
\hline
0.00833(12) &         & -0.0583(13) & 0.0222(6)  & & & 5.1 \\
            & 0.060(6) & -0.104(12) & 0.0225(34) & & & 180 \\
0.00944(32) & -0.0087(24) & -0.0507(23) & 0.0221(4) & & & 2.6 \\
0.00877(18) &             & -0.0656(27) & 0.0220(5) & 0.021(7) & & 3.2 \\
 & 0.116(8) & -0.271(23)  & 0.0213(15) & 0.32(4) & & 34 \\
0.0097(5)   &             & -0.096(14) & 0.0224(5) & & 0.099(37) & 3.5 \\
0.0110(8) & -0.031(11) & -0.010(20) & 0.02223(37) & -0.061(29) & & 2.0 \\
0.0203(11) & -0.282(30) & 0.70(8) & 0.02220(14) & -0.68(7) & -0.71(8) & 0.3 \\
\hline\hline
\end{tabular}
\caption{Parameters for nonlinear fit beyond-TF LDA exchange data for alkali
earths to Eq.~{eq:beyondTF}.  All rows $n\geq 2$ are included in fitting set.  
Blank spaces indicate where a parameter is held at zero.  $\chi^2_{red}$ is
reduced $\chi^2$ measure, defined as the $\chi^2$ error of the fit divided
by the number of independent degrees of freedom in the fit set.}
\label{table:fit2}
\end{table*}

\label{page:end}

\end{document}